\documentclass[runningheads]{llncs}
\pdfoutput=1
\usepackage[pagebackref=true,breaklinks=true,letterpaper=true,colorlinks,bookmarks=false]{hyperref}
\usepackage[]{booktabs}
\usepackage[pdftex]{graphicx}
\usepackage{graphicx}
\usepackage{subcaption}
\usepackage[]{algorithm2e}
\usepackage{amsmath}
\usepackage{color}
\usepackage{mathtools}
\usepackage{tabularx}

\DeclareGraphicsExtensions{.pdf,.jpeg,.png}
\DeclarePairedDelimiter\ceil{\lceil}{\rceil}
\DeclarePairedDelimiter\floor{\lfloor}{\rfloor}

\begin{document}
\pagestyle{headings}
\mainmatter
\title{VideoSet: A Large-Scale Compressed Video Quality Dataset Based on
JND Measurement}
\titlerunning{VideoSet}
\author{
Haiqiang Wang\inst{1}\and
Ioannis Katsavounidis\inst{2}\and
Jiantong Zhou\inst{3}\and
Jeonghoon Park\inst{4}\and
Shawmin Lei\inst{5}\and
Xin Zhou\inst{6}\and
Man-On Pun\inst{7}\and
Xin Jin\inst{8}\and
Ronggang Wang\inst{9}\and
Xu Wang\inst{10}\and
Yun Zhang\inst{11}\and
Jiwu Huang\inst{10}\and
Sam Kwong\inst{12}\and
C.-C. Jay Kuo\inst{1}
}
\authorrunning{Wang et al.}
\institute{
University of Southern California, Los Angeles, California, USA \and
Netflix, Los Gatos, California, USA \and
Huawei Technologies, Shenzhen, China \and
Samsung, DMC R\&D, Seoul, Korea \and
Mediatek, Hsinchu, Taiwan \and
Northwestern Polytechnical University, Xi'an, China \and
Chinese University of Hong Kong (SZ), Shenzhen, China \and
Graduate School at Shenzhen, Tsinghua University, Guangdong, China \and
Peking University Shenzhen Graduate School, Shenzhen, China \and
Shenzhen University, Shenzhen, China \and
Chinese Academy of Sciences, Shenzhen, China \and
City University of Hong Kong, Hong Kong
}
\toctitle{VideoSet}
\tocauthor{VideoSet}
\maketitle

\begin{abstract}
A new methodology to measure coded image/video quality using the
just-noticeable-difference (JND) idea was proposed in
\cite{lin2015experimental}. Several small JND-based image/video quality
datasets were released by the Media Communications Lab at the University
of Southern California in \cite{jin2016jndhvei,mcl_jcv}. In this work,
we present an effort to build a large-scale JND-based coded video
quality dataset. The dataset consists of 220 5-second sequences in four
resolutions (i.e., $1920 \times 1080$, $1280 \times 720$, $960 \times 540$
and $640 \times 360$).  For each of the 880 video clips, we encode it
using the H.264 codec with $QP=1, \cdots, 51$ and measure the first
three JND points with 30+ subjects. The dataset is called the `VideoSet', which is an acronym for `Video Subject Evaluation Test
(SET)'. This work describes the subjective test procedure, detection
and removal of outlying measured data, and the properties of collected
JND data. Finally, the significance and implications of the VideoSet to
future video coding research and standardization efforts are pointed
out. All source/coded video clips as well as measured JND data included
in the VideoSet are available to the public in the IEEE DataPort
\cite{h2h01c-16-full}.
\end{abstract}

\section{Introduction} \label{sec:intro}
Digital video plays an important role in our daily life. About 70\% of
today's Internet traffic is attributed to video, and it will continue to
grow to the 80-90\% range within a couple of years. It is critical to
have a major breakthrough in video coding technology to accommodate the
rapid growth of video traffic. Despite the introduction of a set of
fine-tuned coding tools in the standardization of H.264/AVC and H.265 (or
HEVC), a major breakthrough in video coding technology is needed to meet
the practical demand. To address this problem, we need to examine
limitations of today's video coding methodology.

\begin{figure}[!t]
	\centering
	\begin{subfigure}[b]{1.0\linewidth}
		\centering
		\includegraphics[width=0.75\linewidth]{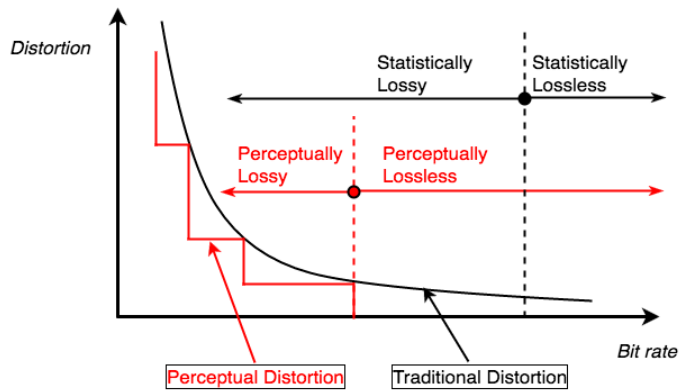}
		\phantomcaption
	\end{subfigure}
	\caption{Comparison between the traditional R-D function and the newly
	observed stair R-D function. The former does not take the nonlinear
	human perception process into account.\label{fig:perceptual_distortion}}
\end{figure}

Today's video coding technology is based on Shannon's source coding
theorem, where a continuous and convex rate-distortion (R-D) function
for a probabilistic source is derived and exploited (see the black curve
in Fig. \ref{fig:perceptual_distortion}). However, humans cannot
perceive small variation in pixel differences. Psychophysics study on
the just-noticeable difference (JND) clearly demonstrated the nonlinear
relation between human perception and physical changes. The traditional
R-D function does not take this nonlinear human perception process into
account. In the context of image/video coding, recent subjective
studies in \cite{lin2015experimental} show that humans can only perceive
discrete-scale distortion levels over a wide range of coding bitrates
(see the red curve in Fig. \ref{fig:perceptual_distortion}).

Without loss of generality, we use H.264 video as an example to explain
it. The quantization parameter (QP) is used to control its quality. The
smaller the QP, the better the quality. Although one can choose a wide
range of QP values, humans can only differentiate a small number of
discrete distortion levels among them. In contrast with the conventional
R-D function, the perceived R-D curve is neither continuous nor convex.
Rather, it is a stair function that contains a couple of jump points,
called just noticeable difference (JND) points. The JND is a
statistical quantity that accounts for the maximum difference
unnoticeable to a human being. Subjective tests for traditional visual
coding and processing were only conducted by very few experts called
golden eyes. This is the worst-case analysis. As the emergence of big
data science and engineering, the worst-case analysis cannot reflect the
statistical behavior of the group-based quality of experience (QoE).
When the subjective test is conducted with respect to a viewer group, it
is more meaningful to study their QoE statistically to yield an
aggregated function.

The measure of coded image/video quality using the JND notion was first
proposed in \cite{lin2015experimental}.  As a follow-up, two small-scale
JND-based image/video quality datasets were released by the Media
Communications Lab at the University of Southern California. They are
the MCL-JCI dataset \cite{jin2016jndhvei} and the MCL-JCV dataset
\cite{mcl_jcv} targeted the JPEG image and the H.264 video,
respectively. To build a large-scale JND-based video quality dataset,
an alliance of academic and industrial organizations was formed and the
subjective test data were acquired in Fall 2016. The resulting dataset
is called the ``VideoSet" -- an acronym for ``Video Subject Evaluation
Test (SET)". The VideoSet consists of 220 5-second sequences in four
resolutions (i.e., $1920 \times 1080$, $1280 \times 720$, $960 \times 540$
and $640 \times 360$). For each of the 880 video clips, we encode it
using the x264 \cite{aimar2005x264} encoder implementation of the H.264 standard with $QP=1, \cdots, 51$ and measure the first
three JND points with 30+ subjects.
All source/coded video clips as well as measured JND data included
in the VideoSet are available to the public in the IEEE DataPort
\cite{h2h01c-16-full}.

The rest of this paper is organized as follows. The source and
compressed video content preparation is discussed in Sec.
\ref{sec:source}. The subjective evaluation procedure is described in
Sec. \ref{sec:subjective}. The outlier detection and removal process is
conducted for JND data post-processing in Sec.
\ref{sec:outlier_detection}. Some general discussion on the VideoSet is
provided in Sec. \ref{sec:experimental_results}. The significance and
implication of the VideoSet to future video coding research and
standardization efforts are pointed out in Sec.  \ref{sec:significance}.
Finally, concluding remarks and future work are given in Sec.
\ref{sec:conclusion}.

\section{Source and Compressed Video Content}\label{sec:source}

We describe both the source and the compressed video content in this section.

\subsection{Source Video} \label{subsub:source_video}

\begin{figure}[!t]
	\centering
	\begin{subfigure}[b]{1.0\linewidth}
		\centering
		\includegraphics[width=1.0\linewidth]{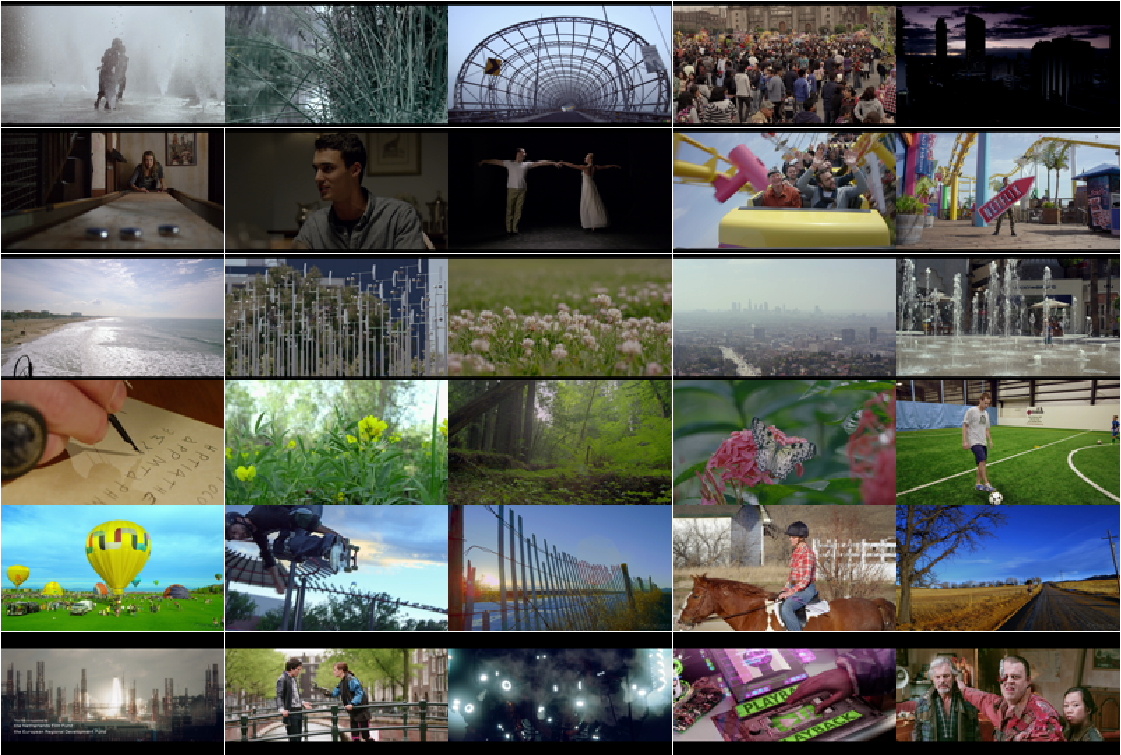}
		\phantomcaption
	\end{subfigure}
\caption{Display of 30 representative thumbnails of video clips from
the VideoSet, where video scenes in the first three rows are from two long
sequences ``El Fuente" and ``Chimera" \cite{CDVL}, those in the fourth and
fifth rows are from the CableLab sequences \cite{Cablelabs}, while those in
the last row are from ``Tears of Steel" \cite{tos}.\label{fig:thumbnails}}
\end{figure}

The VideoSet consists of 220 source video clips, each of which has a
duration of 5 seconds. We show thumbnails images for 30 representative
video clips in Fig \ref{fig:thumbnails}. The source video clips were
collected from publicly available datasets in \cite{CDVL,Cablelabs,tos}.
The original sequences have multiple spatial resolutions ({\em i.e.},
$4096 \times 2160$, $4096 \times 1714$, $3840 \times 2160$), frame
rates ({\em i.e.}, 60, 30, 24) and color formats ({\em i.e.}, YUV444p,
YUV422p, YUV420p). We pay special attention to the selection of these
source video clips to avoid redundancy and enrich diversity of selected
contents.

After content selection, we process each 5-second video clip to ensure
that they are in similar format. Their formats are summarized in Table
\ref{tab:raw_sequences_summary}, where the first column shows the names
of the source video material of longer duration and the second column
indicates the number of video clips selected from each source material.
The third, fourth and fifth columns describe the frame rate, the spatial
resolution and the pixel format, respectively. They are further
explained below.

\begin{itemize}
\item {\bf Frame Rate.} The frame rate affects the perceptual quality of
certain contents significantly \cite{ou2014q}. Contents of a higher
frame rate ({\em e.g.} 60fps) demand a more powerful CPU and a larger
memory to avoid impairments in playback. For this reason, if the
original frame rate is 60fps, we convert it from 60fps to 30fps to
ensure smooth playback in a typical environment. If the original frame
rate is not greater than 30fps, no frame rate conversion is needed.
\item {\bf Spatial Resolution.} The aspect ratio of most commonly used
display resolutions for web users is $16:9$. For inconsistent aspect
ratios, we scale them to $16:9$ by padding black horizontal bars above
and below the active video window. As a result, all video clips are of
the same spatial resolution -- $3840 \times 2160$.
\item {\bf Pixel Format.} We down-sample the trimmed spatial resolution
$3840 \times 2160$ (2160p) to four lower resolutions. They are: $1920
\times 1080$ (1080p), $1280 \times 720$ (720p), $ 960 \times 540$ (540p)
and $640 \times 360$ (360p) for the subjective test in building the
VideoSet. In the spatial down-sampling process, the lanczos
interpolation \cite{lanczos} is used to keep a good compromise between
low and high frequencies components. Also, the $4:2:0$ chroma sampling
is adopted for maximum compatibility.
\end{itemize}

It is worthwhile to point out that 1080p and 720p are two most dominant
video formats on the web nowadays while 540p and 360p are included to
capture the viewing experience on tablets or mobile phones. After the
above-mentioned processing, we obtain 880 uncompressed sequences in total.

\begin{table}[!t]
\centering
\caption{Summarization of source video formats in the VideoSet.}\label{tab:raw_sequences_summary}
\vspace{3mm}
\resizebox{\columnwidth}{!}{
\begin{tabular}{*{8}{c}}
\toprule
 & & \multicolumn{2}{c}{Frame rate} & \multicolumn{2}{c}{Spatial resolution}
& \multicolumn{2}{c}{Pixel format} \\
\cmidrule(r){3-4} \cmidrule(r){5-6} \cmidrule(r){7-8}
Source              & Selected & Original & Trimmed & Original             & Trimmed              & Original & Trimmed \\
\midrule
El Fuente           & 31       & 60       &  30     & $4096 \times 2160$   & $3840 \times 2160$   &  YUV444p & YUV420p \\
Chimera             & 59       & 30       &  30     & $4096 \times 2160$   & $3840 \times 2160$   &  YUV422p & YUV420p \\
Ancient Thought     & 11       & 24       &  24     & $3840 \times 2160$   & $3840 \times 2160$   &  YUV422p & YUV420p \\
Eldorado            & 14       & 24       &  24     & $3840 \times 2160$   & $3840 \times 2160$   &  YUV422p & YUV420p \\
Indoor Soccer       & 5        & 24       &  24     & $3840 \times 2160$   & $3840 \times 2160$   &  YUV422p & YUV420p \\
Life Untouched      & 15       & 60       &  30     & $3840 \times 2160$   & $3840 \times 2160$   &  YUV444p & YUV420p \\
Lifting Off         & 13       & 24       &  24     & $3840 \times 2160$   & $3840 \times 2160$   &  YUV422p & YUV420p \\
Moment of Intensity & 10       & 60       &  30     & $3840 \times 2160$   & $3840 \times 2160$   &  YUV422p & YUV420p \\
Skateboarding       & 9        & 24       &  24     & $3840 \times 2160$   & $3840 \times 2160$   &  YUV422p & YUV420p \\
Unspoken Friend     & 13       & 24       &  24     & $3840 \times 2160$   & $3840 \times 2160$   &  YUV422p & YUV420p \\
Tears of Steel      & 40       & 24       &  24     & $4096 \times 1714$   & $3840 \times 2160$   &  YUV420p & YUV420p \\
\bottomrule
\end{tabular}}
\end{table}

\subsection{Video Encoding} \label{sub:sequence_encoding}

We use the H.264/AVC \cite{aimar2005x264} high profile to encode each of the 880
sequences, and choose the constant quantization parameter (CQP) as the
primary bit rate control method. The adaptive QP adjustment is reduced
to the minimum amount since our primary goal is to understand a direct
relationship between the quantization parameter and perceptual quality.
The encoding recipe is included in the read-me file of the released
dataset.

The QP values under our close inspection are between $[8, 47]$. It is
unlikely to observe any perceptual difference between the source and
coded clips with a QP value smaller than $8$. Furthermore, coded video
clips with a QP value larger than $47$ will not be able to offer
acceptable quality. On the other hand, it is ideal to examine the full
QP range; namely, $[0, 51]$, in the subjective test since the JND
measure is dependent on the anchor video that serves as a fixed
reference.

To find a practical solution, we adopt the following modified scheme.
The reference is losslessly encoded and referred to as $QP=0$. We use
the source $QP=0$ to substitute all sequences with a QP value smaller
than 8. Similarly, sequences with a QP larger value than 47 are
substituted by that with $QP=47$. The modification has no influence on
the subjective test result. This will become transparent when we
describe the JND search procedure in Sec.
\ref{sub:jnd_search_procedure}. By including the source and all coded
video clips, there are $220 \times 4 \times 52 = 45,760$ video clips in
the VideoSet.

\section{Subjective Test Environment and Procedure}\label{sec:subjective}

The subjective test environment and procedure are described in detail
in this section.

\subsection{Subjective Test Environment}

\begin{figure}[!t]
\centering
	\begin{subfigure}[b]{0.48\linewidth}
		\centering
		\includegraphics[width=1.0\linewidth]{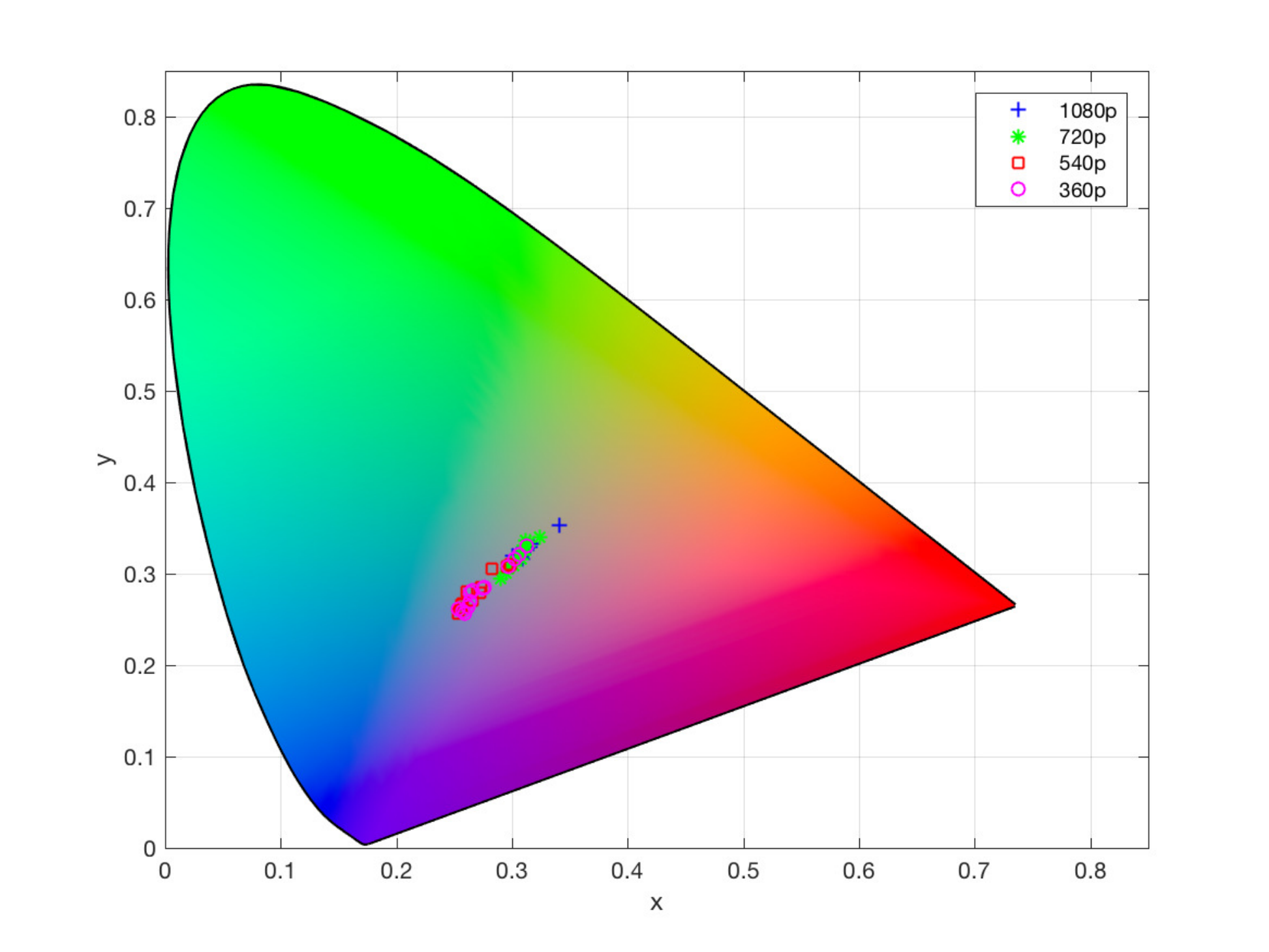}
		\caption{\label{fig:cie_1931_xy}}
	\end{subfigure}
	\begin{subfigure}[b]{0.48\linewidth}
		\centering
		\includegraphics[width=1.0\linewidth]{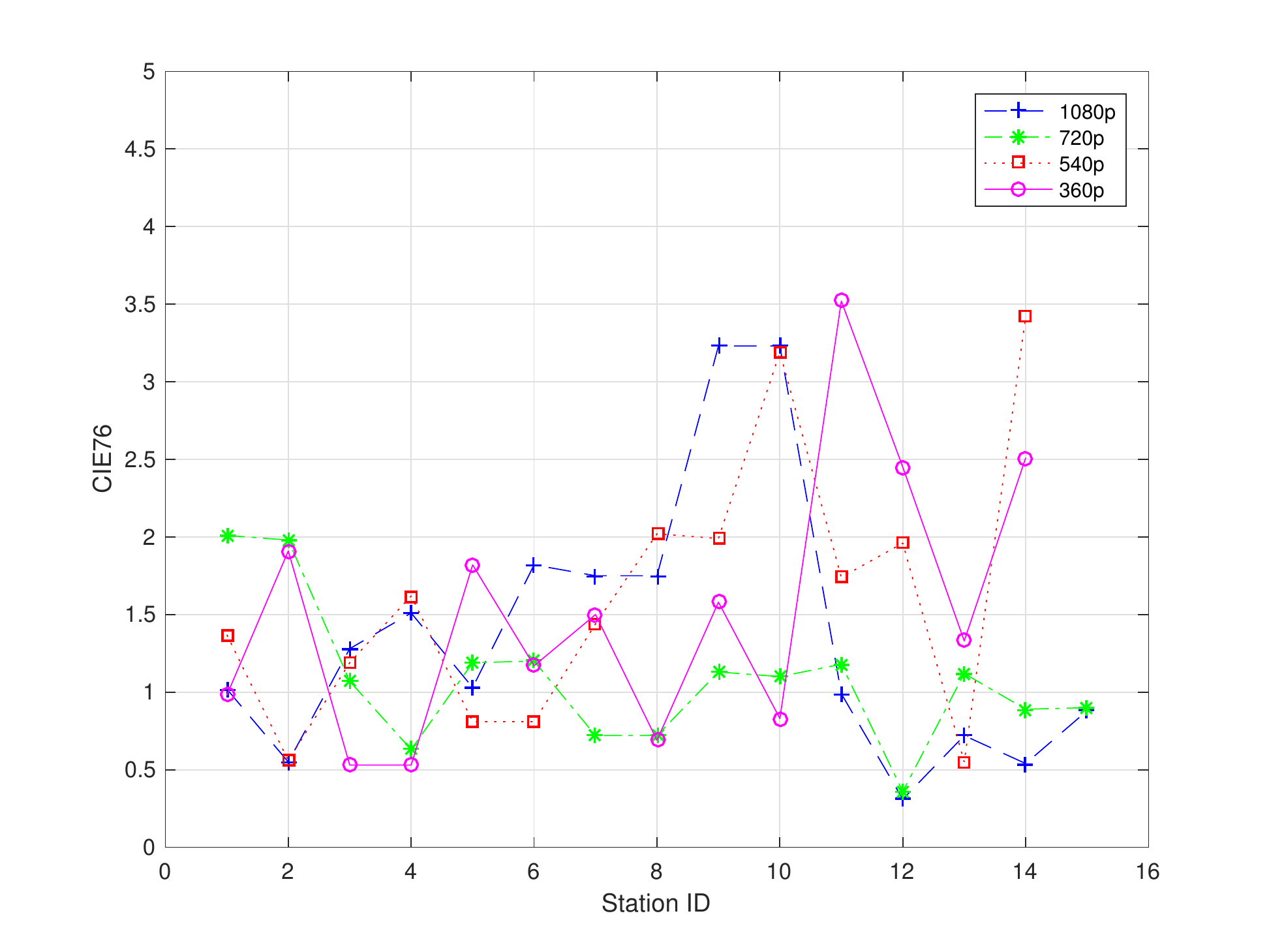}
		\caption{\label{fig:delta_E}}
	\end{subfigure}
	\begin{subfigure}[b]{0.48\linewidth}
		\centering
		\includegraphics[width=1.0\linewidth]{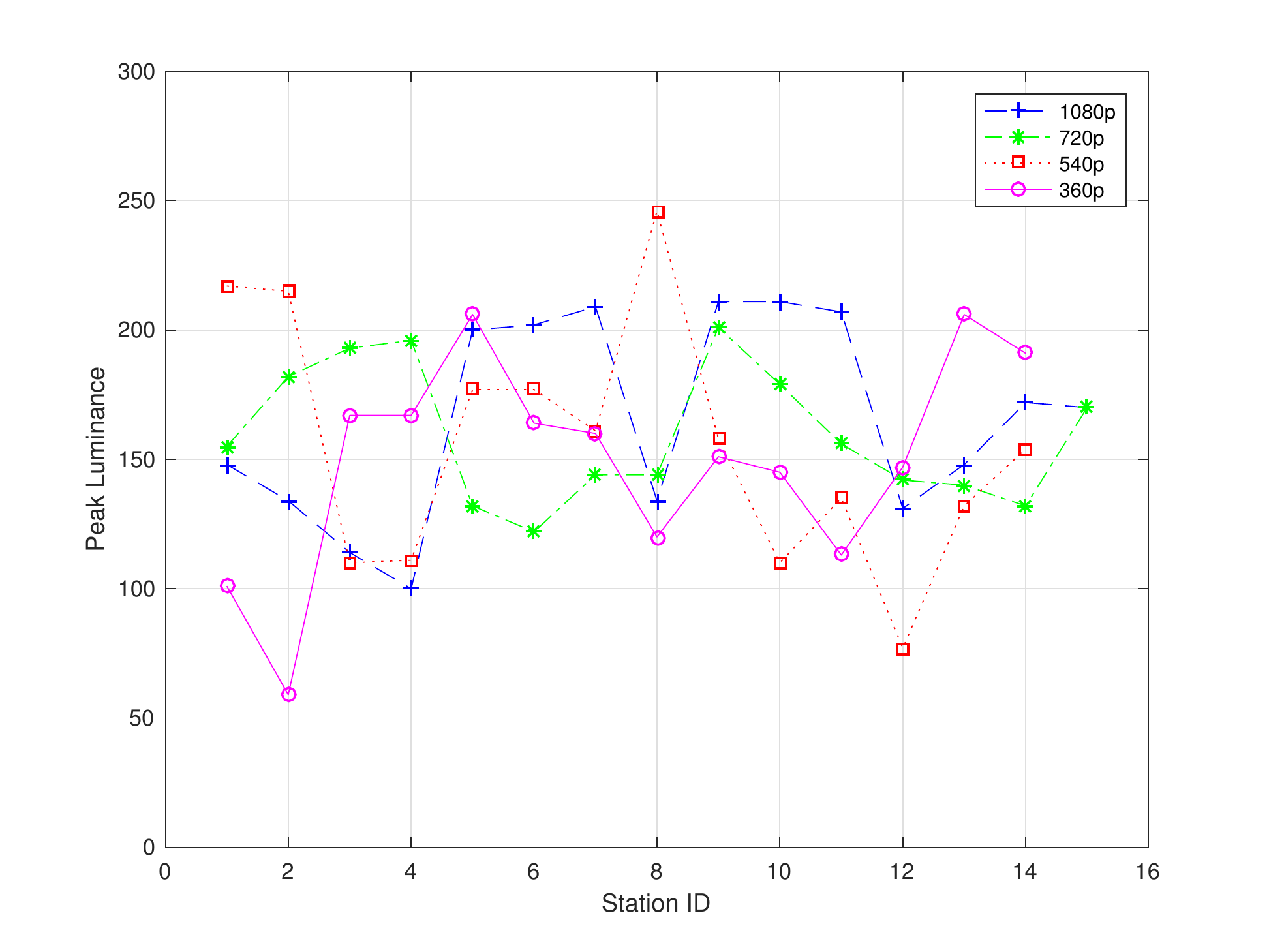}
		\caption{\label{fig:peak_lumi}}
	\end{subfigure}
	\begin{subfigure}[b]{0.48\linewidth}
		\centering
		\includegraphics[width=1.0\linewidth]{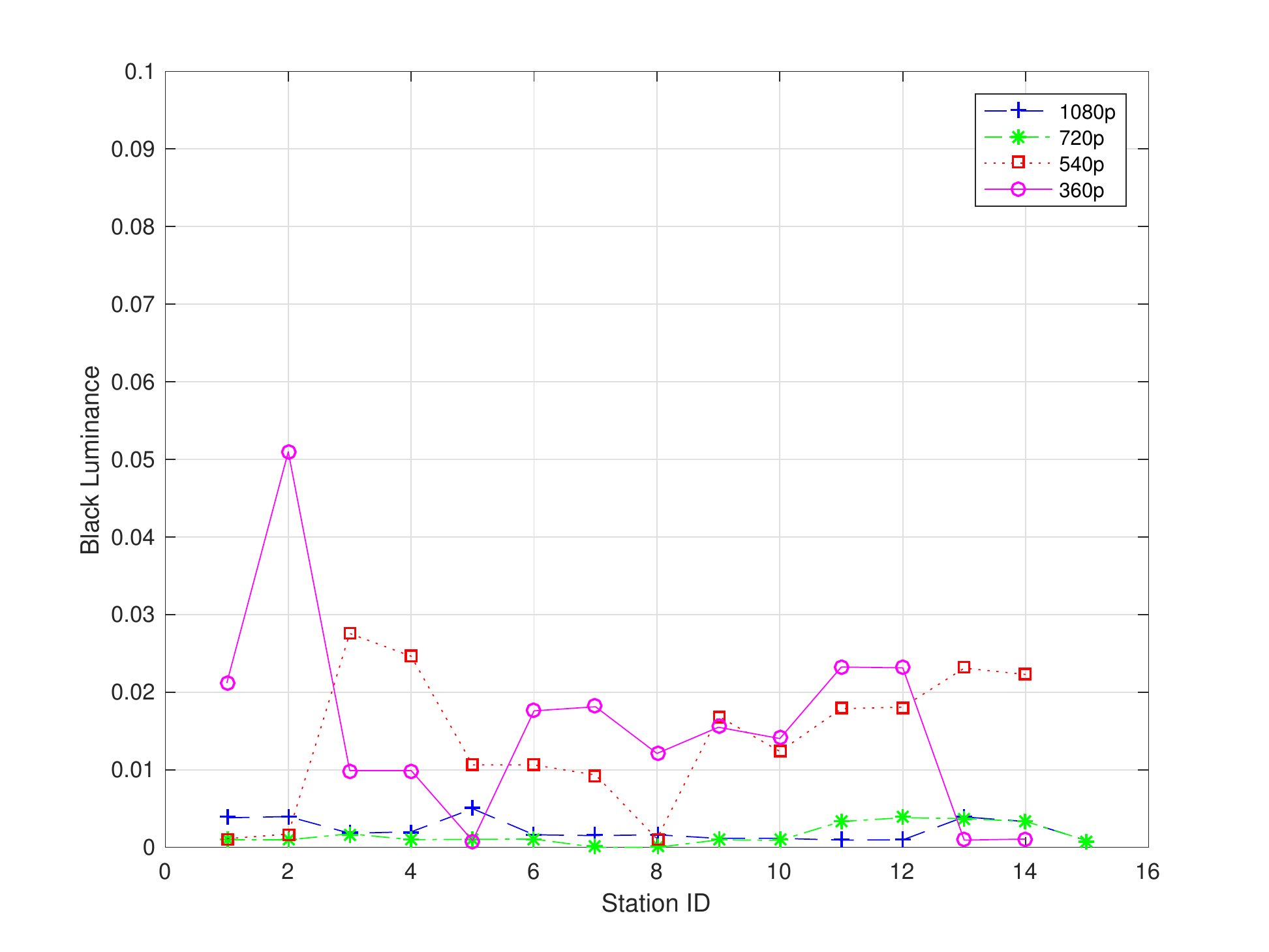}
		\caption{\label{fig:black_lumi}}
	\end{subfigure}
\caption{Results of monitor profiling: (a) chromaticity of white color
in the CIE 1931 color space, (b) the color difference between a specific
monitor and the standard, where $\Delta E \approx 2.3$ corresponds to a
JND \cite{sharma2002digital}, (c) the peak luminance of the screen,
and (d) the luminance ratio of the screen ({\em i.e.,} the luminance of
the black level to the peak white.)} \label{fig:monitor_profiling}
\end{figure}

The subjective test was conducted in six universities in the city of
Shenzhen in China. There were 58 stations dedicated to the subjective
test. Each station offered a controlled non-distracting laboratory
environment. The viewing distance was set as recommended in ITU-R
BT.2022. The background chromaticity and luminance were set up as an
environment of a common office/laboratory. We did not conduct monitor
calibration among different test stations, yet the monitors were
adjusted to a comfortable setting to test subjects. On one hand, the
uncalibrated monitors provided a natural platform to capture the
practical viewing experience in our daily life. On the other hand, the
monitors used in the subjective test were profiled for completeness.
Monitor profiling results are given in Fig. \ref{fig:monitor_profiling}
and summarized in Table \ref{tab:environment_check}. As shown in Table
\ref{tab:environment_check}, most stations comply with ITU
recommendations.

We indexed each video clip with a content ID and a resolution ID and
partitioned 880 video clips into 58 packages. Each package contains 14
or 15 sequence sets of a content/resolution ID pair, and each sequence
set contains one source video clip and its all coded video clips. One
subject can complete one JND point search for one package in one test
session. The duration of one test session was around 35 minutes with a
5-minute break in the middle. Video sequences were displayed in their
native resolution without scaling on the monitor. The color of the
inactive screen was set to light gray.

We randomly recruited around 800 students to participate in the
subjective test. A brief training session was given to each subject
before a test session starts. In the training session, we used different
video clips to show quality degradation of coded video contents. The
scenario of our intended application; namely, the daily video streaming
experience, was explained. Any question from the subject about the
subjective test was also answered.

\begin{table}[!t]
\centering
\caption{Summary on test stations and monitor profiling results.  The
peak luminance and the black luminance columns show the numbers of
stations that meet ITU-R BT.1788 in the corresponding metrics,
respectively. The color difference column indicates the number of
stations that has the $\Delta E$ value smaller than a JND threshold. The
$H$ value indicates the active picture height.\label{tab:environment_check}}
\vspace{3mm}
\resizebox{\columnwidth}{!}{
\begin{tabular}{*{6}{c}} \toprule
Resolution          & Station Number & Peak Luminance ($cd/m^2$) &
Black Luminance & Color Difference & Viewing Distance ($H$)\\  \midrule
1080p               & $15$           & $15$      &  $15$     & $13$       &  $3.2$ \\
720p                & $15$           & $15$      &  $15$     & $15$       &  $4.8$ \\
540p                & $14$           & $14$      &  $14$     & $13$       &  $6.4$ \\
360p                & $14$           & $13$      &  $14$     & $11$       &  $7$ \\
\bottomrule
\end{tabular}}
\end{table}

\subsection{Subjective Test Procedure} \label{sub:jnd_search_procedure}

In the subjective test, each subject compares the quality of two clips
displayed one after another, and determines whether these two sequences
are noticeably different or not. The subject should choose either `YES'
or `NO' to proceed. The subject has an option to ask to play the two
sequences one more time. The comparison pair is updated based on the
response.

One aggressive binary search procedure was described in \cite{mcl_jcv}
to speed up the JND search process. At the first comparison, the
procedure asked a subject whether there would be any noticeable
difference between $QP=0$ and $QP=25$. If a subject made an unconfident
decision of `YES' at the first comparison, the test procedure would
exclude interval $QP=[26, 51]$ in the next comparison. Although the
subjects selects `Noticeable Difference' in all comparisons afterwards,
the final JND location would stay at $QP=25$. It could not belong to
$QP=[26, 51]$ any longer. A similar problem arose if a subject made an
unconfident decision of `NO' at the first comparison.

To fix this problem, we adopt a more robust binary search procedure in
our current subjective test. Instead of eliminating the entire left or
right half interval, only one quarter of the original interval at the
farthest location with respect to the region of interest is dropped in
the new test procedure. Thus, if a subject made an unconfident decision
of `YES' at the first comparison, the test procedure will remove
interval $QP=[39, 51]$ so that the updated interval is $QP=[0,38]$.  The
new binary search procedure allows a buffer even if a wrong decision is
made. The comparison points may oscillate around the final JND position
but still converge to it. The new binary search procedure is proved to
be more robust than the previous binary search procedure at the cost of
a slightly increased number of comparisons ({\em i.e.}, from 6
comparisons in the previous procedure to 8 comparisons in the new
procedure).

Let $x_{n} \in{[0,51]}$ be the QP used to encode a source sequence.  We
use $x_{s}$ and $x_{e}$ as the start and the end QP values of a search
interval, $[x_{s}, x_{e}]$, at a certain round.  Since $x_{s} < x_{e}$,
the quality of the coded video clip with $QP=x_s$ is better than that
with $QP=x_e$. We use $x_{a}$ to denote the QP value of the anchor video
clip. It is fixed in the entire binary search procedure until the JND
point is found. The QP value, $x_{c}$, of the comparison video is
updated within $[x_{s},x_{e}]$. One round of the binary search
procedure is described in Algorithm \ref{algo}.

\begin{algorithm}[!t]
    \KwData{QP range $[x_{s},x_{e}]$}
    \KwResult{JND location $x_{n}$}
    $x_{a}=x_{s}$\;
    $x_{l}=x_{s}$\;
    $x_{r}=x_{e}$\;
    flag = true\;
    \While{flag}{
        \eIf{ $x_{a}$ and $x_{c}$ have quality difference}{
            $x_{n}=x_{c}$\;
            \eIf{$x_{c}-x_{l} \leq 1$}{
                flag=false\;
            }
            {
            $x_{r}=\floor*{(x_{l}+3*x_{r}/4}$\;
            $x_{c}=\floor*{(x_{l}+x_{r})/2}$ \;
            }
        }
        {
            \eIf{$x_{r}-x_{c} \leq 1$}{
                flag=false\;
            }
            {
            $x_{l}=\ceil*{(3*x_{l}+x_{r})/4}$\;
            $x_{c}=\ceil*{(x_{l}+x_{r})/2}$ \;
            }
        }
    }
\caption{One round of the JND search procedure.}\label{algo}
\end{algorithm}

The global JND search algorithm is stated below.
\begin{itemize}
\item Initialization. We set $x_{s}=0$ and $x_{e}=51$.
\item Search range update. If $x_{a}$ and $x_{c}$ exhibit a noticeable
quality difference, update $x_{r}$ to the third quartile of the range.
Otherwise, update $x_{l}$ to the first quartile of the range.  The
ceiling and the floor integer-rounded operations, denoted by $\floor*{*}$
and $\ceil*{*}$, are used in the update process as shown in the
Algorithm of the one round of the JND search procedure.
\item Comparison video update. The QP value of the comparison video clip
is set to the middle point of the range under evaluation with the
integer-rounded operation.
\item Termination. There are two termination cases. First, if
$x_{c}-x_{l} \leq 1$ and the comparison result is `Noticeable
Difference', then search process is terminated and $x_{c}$ is set to the
JND point. Second, if $x_{r}-x_{c} \leq 1$ and the comparison result is
`Unnoticeable Difference', the process is terminated and the JND is the
latest $x_{c}$ when the comparison result was `Noticeable Difference'.
\end{itemize}

The JND location depends on the characteristics of the underlying video
content, the visual discriminant power of a subject and the viewing
environment.  Each JND point can be modeled as a random variable with
respect to a group of test subjects.  We search and report three JND
points for each video clip in the VideoSet.  It will be argued in Sec.
\ref{sec:experimental_results} that the acquisition of three JND values
are sufficient for practical applications.

For a coded video clip set, the same anchor video is used for all test
subjects. The anchor video selection procedure is given below. We plot
the histogram of the current JND point collected from all subjects and
then set the QP value at its first quartile as the anchor video in the
search of the next JND point. For this QP value, $75\%$ of test subjects
cannot notice a difference. We select this value rather than the median
value, where $50\%$ of test subjects cannot see a difference, so as to
set up a higher bar for the next JND point. The first JND point search
is conducted for QP belonging to $[0,51]$. Let $x_{N}$ be the QP value
of the $N^{th}$ JND point for a given sequence. The QP search range for
$(N+1)^{th}$ JND is $[x_{N}, 51]$.

\section{JND Data Post-Processing via Outlier Removal}\label{sec:outlier_detection}

Outliers refer to observations that are significantly different from the
majority of other observations. The notation applies to both test
subjects and collected samples. In practice, outliers should be
eliminated to allow more reliable conclusion. For JND data
post-processing, we adopt outlier detection and removal based on the
individual subject and collected JND samples. They are described below.

\subsection{Unreliable Subjects}

As described in Sec. \ref{sub:sequence_encoding}, video clips are
encoded with $QP=[8,47]$ while $QP=0$ denotes the source video without
any quality degradation. The QP range is further extended to $[0,51]$ by
substituting video of $QP=[1,7]$ with video of $QP=0$, and video of
$QP=[48,51]$ with video of $QP=47$. With this substitution, the video
for $QP=[1,7]$ is actually lossless, and no JND point should lie in this
range. If a JND sample of a subject comes to this interval, the subject
is treated as an outlier. All collected samples from this subject are
removed.

The ITU-R BT 1788 document provides a statistical procedure on subject
screening. It examines score consistency of a subject against all
subjects in a test session, where the scores typically range from 1 to 5
denoting from the poorest to the best quality levels. This is achieved
by evaluating the correlation coefficient between the scores of a
particular subject with the mean scores of all subjects for the whole
test session, where the Pearson correlation or the Spearman rank
correlation is compared against a pre-selected threshold. However, this
procedure does not apply to the collected JND data properly since our
JND data is the QP value of the coded video that meets the just
noticeable difference criterion.

Alternatively, we adopt the z-scores consistency check. Let
$\boldsymbol{x}_n^{m}$ be the samples obtained from subject $m$ on a
video sequence set with video index $n$, where $m = 1, 2, \dots, M$ and
$n = 1, 2, \dots, N$. For subject $m$, we can form a vector of his/her
associated samples as
\begin{equation}
\boldsymbol{x}^{m} = (x_{1}^{m}, x_{2}^{m}, \dots, x_{N}^{m}).
\end{equation}
Its mean and standard deviation (SD) vectors against all subjects can be written as
\begin{eqnarray}
\boldsymbol{\mu} & = & (\mu_{1}, \mu_{2}, \dots, \mu_{N}),
\quad \mu_{n}=\frac{1}{M}\sum_{m=1}^{M} x^{m}_{n}, \\
\boldsymbol{\sigma} & = & (\sigma_{1}, \sigma_{2}, \dots, \sigma_{N}),
\quad \sigma_{n}=\sqrt{\frac{1}{M-1}\sum_{m=1}^{M} (x^{m}_{n}-\mu_{n})^2}.
\end{eqnarray}
Then, the z-scores vector of subject $m$ is defined as
\begin{equation}
\boldsymbol{z}^{m} = (z^{m}_{1}, z^{m}_{2}, \dots, z^{m}_{N}),
\quad z^{m}_{n} = \frac{x^{m}_{n}-\mu_{n}}{\sigma_{n}}.
\end{equation}
The quantity, $z^{m}_{n}$, indicates the distance between the raw score
and the population mean in the SD unit for subject $m$ and video clip
$n$. The dispersion of the z-score vector shows consistency of an
individual subject with respect to the majority.

\begin{figure}[!ht]
\centering
	\begin{subfigure}[b]{0.68\linewidth}
		\centering
		\includegraphics[width=1.0\linewidth]{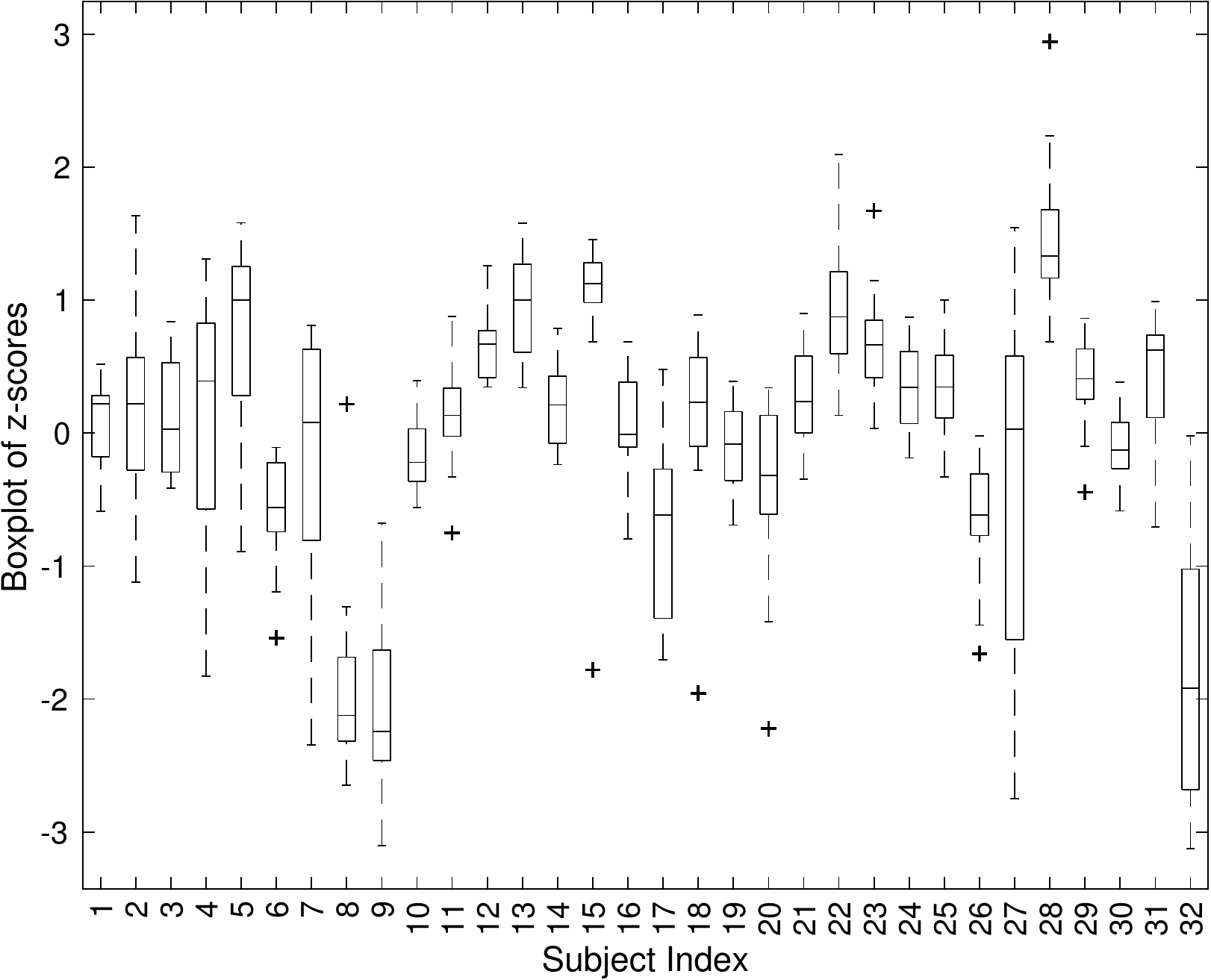}
		\caption{\label{fig:1080_12_raw_boxplot_z_scores}}
	\end{subfigure}

	\begin{subfigure}[b]{0.8\linewidth}
		\centering
		\includegraphics[width=1.0\linewidth]{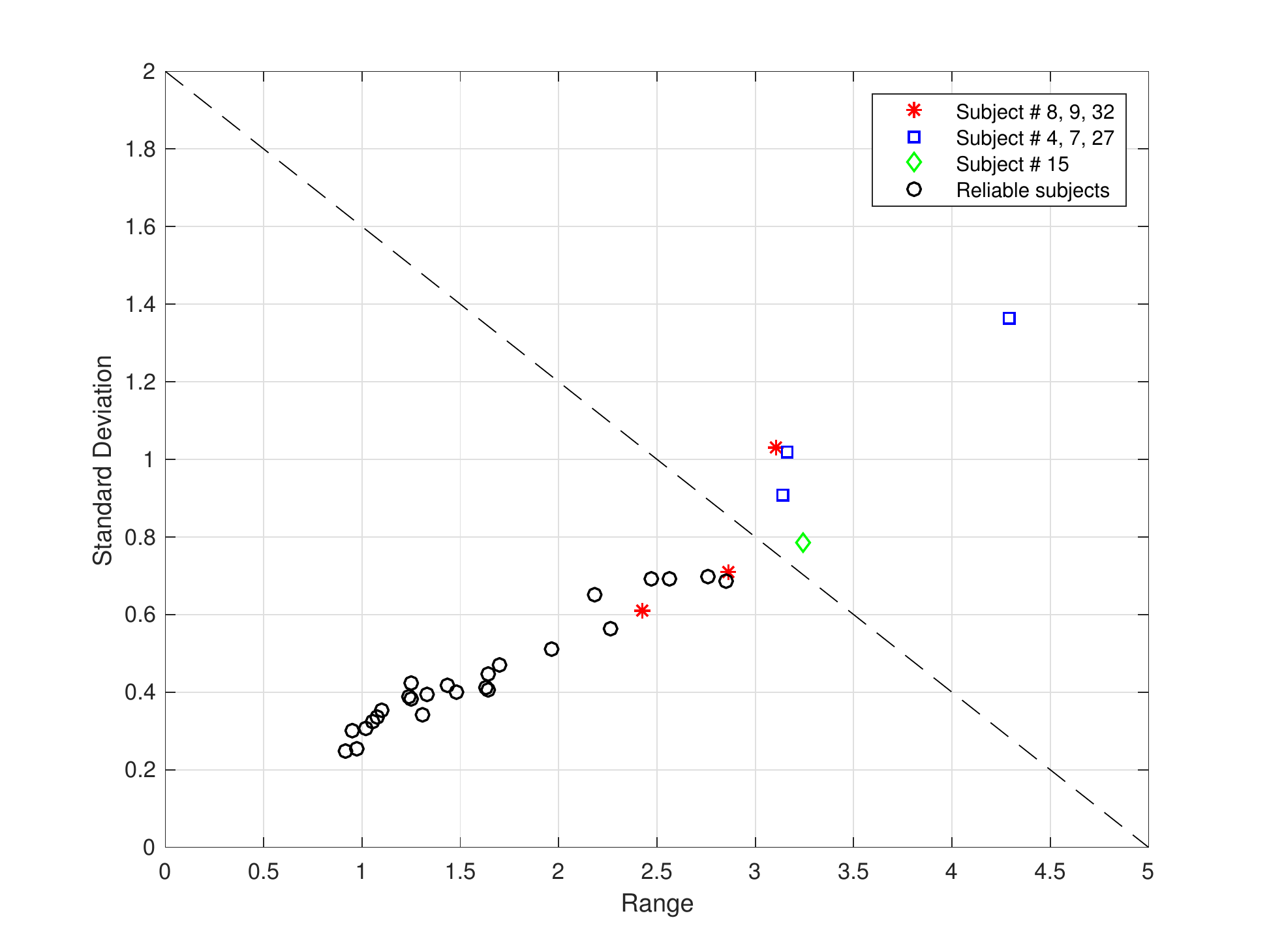}
		\caption{\label{fig:1080_12_dispersion}}
	\end{subfigure}
\caption{Illustration of unreliable subject detection and removal: (a)
the boxplot of z-scores and (b) the dispersion plot of all subjects
participating in the same test session, where subjects \#4, \#7, \#8,
\#9, \#32 and \#37 are detected as outliers. Subject \#15 is kept after
removing one sample.} \label{fig:unreliable_subjects}
\end{figure}

Both the range and the SD of the z-score vector, $\boldsymbol{z}^{m}$,
are used as the dispersion metrics. They are defined as
\begin{equation}
R = \max(\boldsymbol{z}^{m}) - \min(\boldsymbol{z}^{m}), \mbox{   and   }
D = std(\boldsymbol{z}^{m}),
\end{equation}
respectively. A larger dispersion indicates that the corresponding
subject gives inconsistent evaluation results in the test. A subject is
identified as an outlier if the associated range and SD values are both
large. An example is shown in Fig. \ref{fig:unreliable_subjects}. We
provide the boxplot of z-scores for all 32 subjects in Fig.
\ref{fig:1080_12_raw_boxplot_z_scores} and the corresponding dispersion
plot in Fig. \ref{fig:1080_12_dispersion}. The horizontal and the
vertical axes of Fig. \ref{fig:1080_12_dispersion} are the range and the
SD metrics, respectively.

For this particular test example, subjects \#8, \#9 and \#32 are
detected as outliers because some of their JND samples have $QP=1$.
Subjects \#4, \#7 and \#27 are removed since their range and SD are both
large. For subject \#15, the SD value is small yet the range is large
due to one sample. We remove that sample and keep others.

\subsection{Outlying Samples}

Besides unreliable subjects, we consider outlying samples for a
given test content. This may be caused by the impact of the unique
characteristics of different video contents on the perceived quality of
an individual. Here, we use the Grubbs' test \cite{grubbs1950sample} to
detect and remove outliers. It detects one outlier at a
time. If one sample is declared as an outlier, it is removed from the
dataset, and the test is repeated until no outliers are detected.

We use $\boldsymbol{s} = (s_{1}, s_{2}, \dots, s_{N})$ to denote a set
of raw samples collected for one test sequence. The test statistics is
the largest absolute deviation of a sample from the sample mean in the
SD unit. Mathematically, the test statistics can be expressed as
\begin{equation}
G=\frac{\max \limits_{i=1,\dots,N}\left|s_{i}-\bar{\boldsymbol{s}}
\right|}{\sigma_{\boldsymbol{s}}}.
\end{equation}
At a given significant level denoted by $\alpha$, a sample is declared
as an outlier if
\begin{equation}
G>\frac{N-1}{N}\sqrt{\frac{t_{\alpha/(2N),N-2}^{2}}{N-2+t_{\alpha/(2N),N-2}^{2}}},
\end{equation}
where $t_{\alpha/(2N),N-2}^{2}$ is the upper critical value of the
t-distribution with $N-2$ degrees of freedom.  In our subjective test,
the sample size is around $N=30$ after removing unreliable subjects and
outlying samples. We set the significance level at $\alpha=0.05$ as
a common scientific practice. Then, a sample is identified as an outlier
if its distance to the sample mean is larger than the $2.9085$ SD unit.

\subsection{Normality of Post-processed JND Samples}\label{sec:analysis_on_jnd_samples}

\begin{table}[!t]
\centering
\caption{The percentages of JND samples that pass normality test,
where the total sequence number is 220.}\label{tab:normality_check}
\vspace{3mm}
\begin{tabular}{*{4}{c}}
\toprule
Resolution          & The first JND & The second JND & The third JND\\ \midrule
    1080p           & $95.9\%$       & $95.9\%$       &  $93.2\%$     \\
    720p            & $94.1\%$       & $98.2\%$       &  $95.9\%$     \\
    540p            & $94.5\%$       & $97.7\%$       &  $96.4\%$     \\
    360p            & $95.9\%$       & $97.7\%$       &  $95.5\%$     \\
    \bottomrule
\end{tabular}
\end{table}

Each JND point is a random variable. We would like to check whether it
can be approximated by a Gaussian random variable \cite{mcl_jcv} after
outlier removal. The $\beta_{2}$ test was suggested in ITU-R BT.500 to
test whether a collected set of samples is normal or not. It calculates
the kurtosis coefficient of the data samples and asserts that the
distribution is Gaussian if the kurtosis is between 2 and 4.

Here, we adopt the Jarque-Bera test \cite{jarque1987test} to conduct the
normality test. It is a two-sided goodness-of-fit test for normality of
observations with unknown parameters. Its test statistic is defined as
\begin{equation} \label{eq:jb_test}
JB = \frac{n}{6}(s^{2} + \frac{(k-3)^2}{4}),
\end{equation}
where $n$ is the sample size, $s$ is the sample skewness and $k$ is the
sample kurtosis. The test rejects the null hypothesis if the statistic
$JB$ in Eq. (\ref{eq:jb_test}) is larger than the precomputed critical
value at a given significance level, $\alpha$. This critical value can
be interpreted as the probability of rejecting the null hypothesis given
that it is true.

We show the percentage of sequences passing normality test in Table
\ref{tab:normality_check}. It is clear from the table that a great
majority of JND points do follow the Gaussian distribution after the
post-processing procedure.

\begin{figure}[!ht]
\centering
	\begin{subfigure}[b]{0.75\linewidth}
	\centering
	\includegraphics[width=1.0\linewidth]{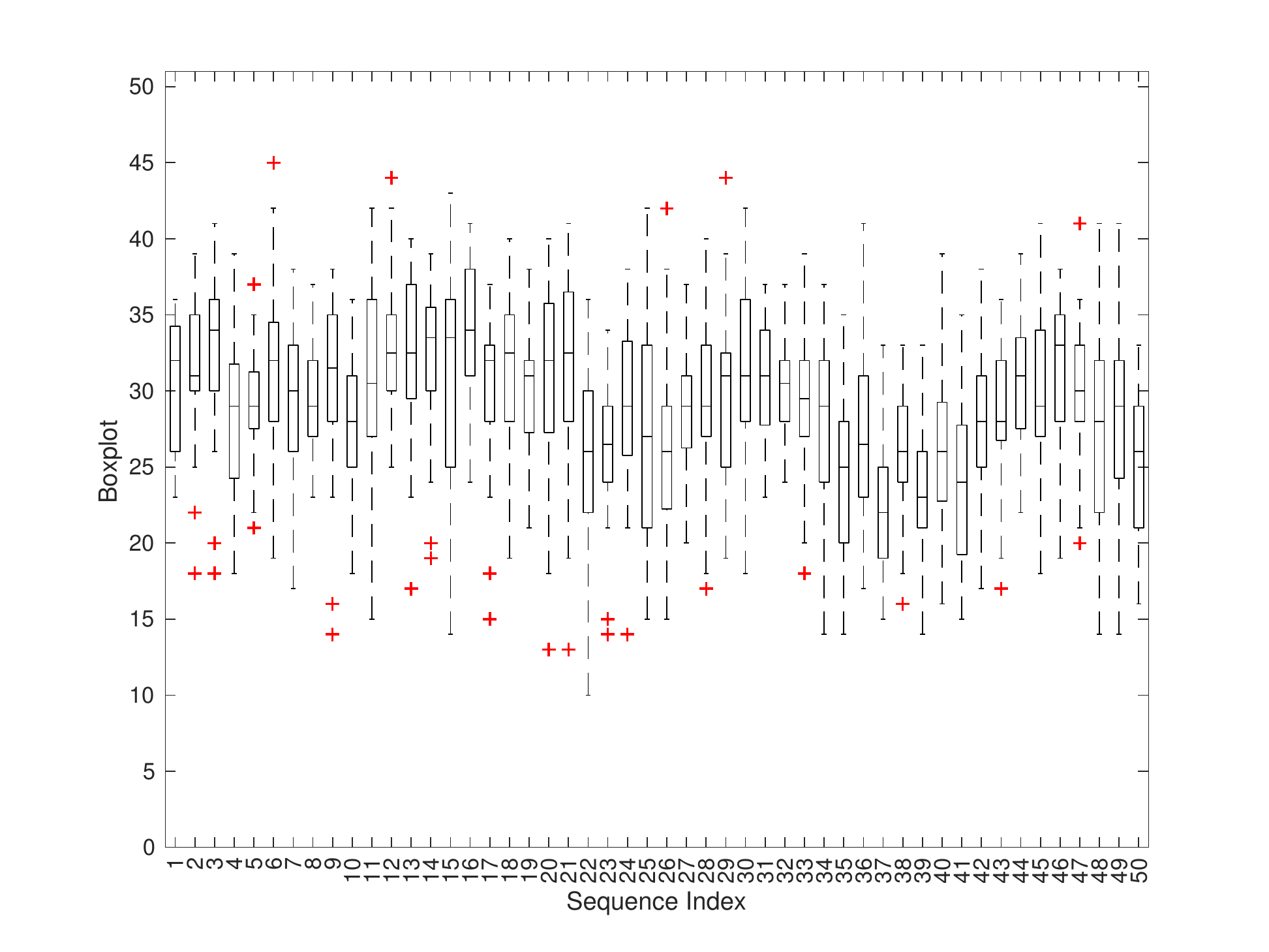}
	\caption{\label{fig:boxplot_1st_1080p_part_1}}
	\end{subfigure}

	\begin{subfigure}[b]{0.48\linewidth}
	\centering
	\includegraphics[width=1.0\linewidth]{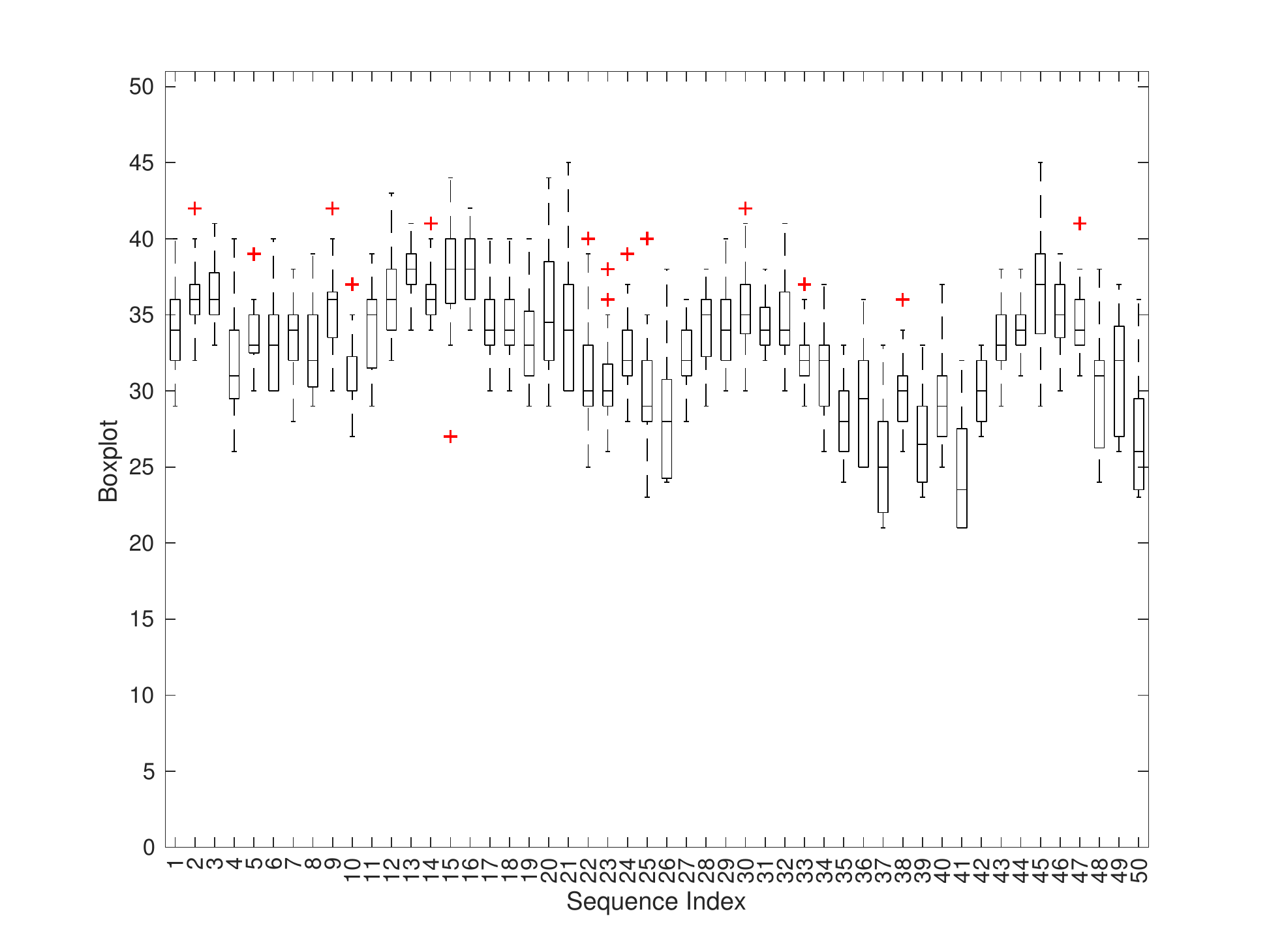}
	\caption{\label{fig:boxplot_2nd_1080p_part_1}}
	\end{subfigure}
	\quad
	\begin{subfigure}[b]{0.45\linewidth}
	\centering
	\includegraphics[width=1.0\linewidth]{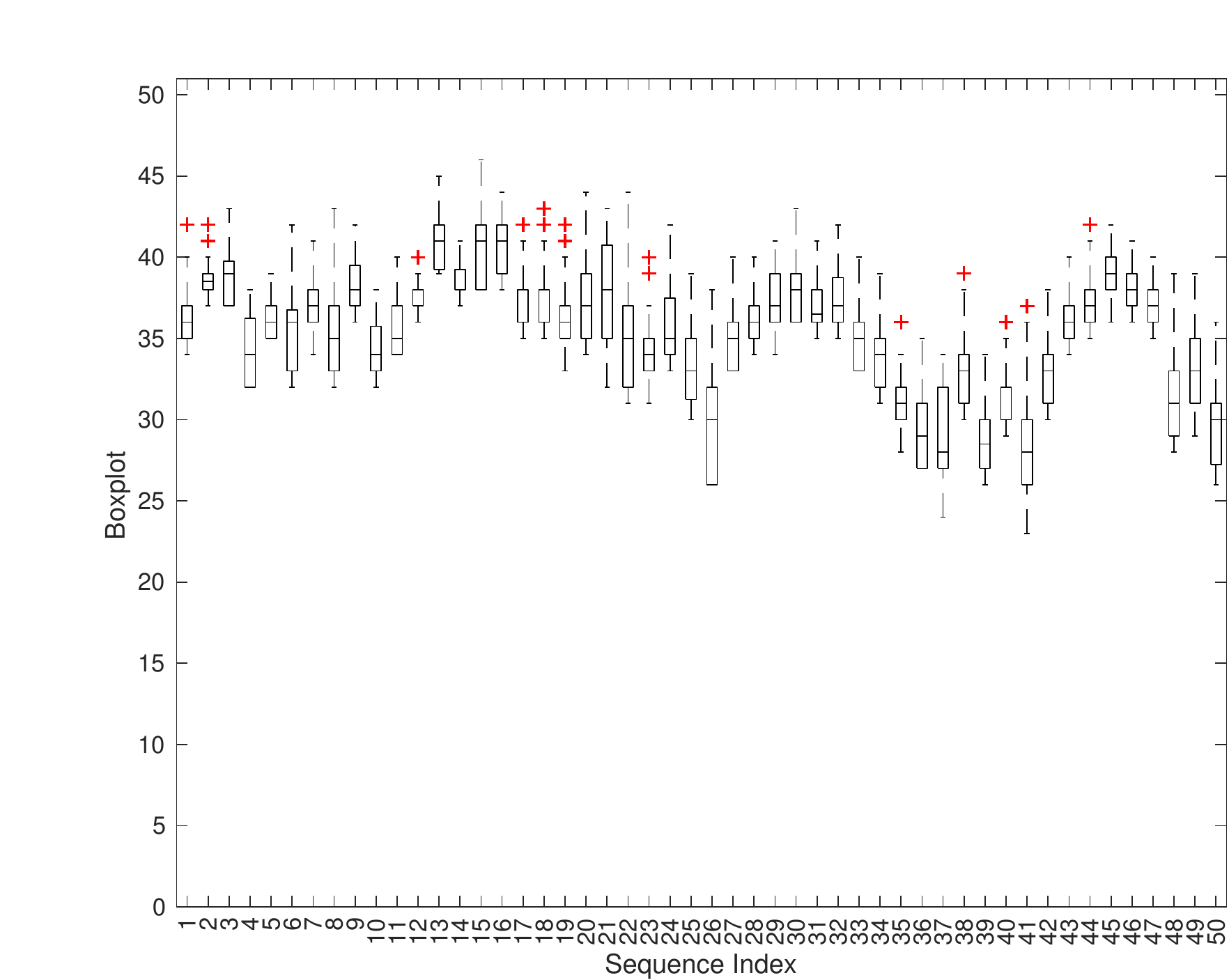}
	\caption{\label{fig:boxplot_3rd_1080p_part_1}}
	\end{subfigure}
\caption{The boxplot of JND samples of the first 50 sequences with
resolution 1080p: (a) the first JND, (b) the second JND, and (c) the
third JND. The bottom, the center and the top edges of the box indicate
the first, the second and the third quartiles, respectively. The bottom
and top whiskers correspond to an interval ranging between $[-2.7\sigma,
2.7\sigma]$, which covers $99.3\%$ of all collected samples.}
\label{fig:boxplot_1080p}
\end{figure}

\section{Discussion}\label{sec:experimental_results}

We show the JND distribution of the first 50 sequences (out of 220
sequences in total) with resolution 1080p in Fig.
\ref{fig:boxplot_1080p}. The figure includes three sub-figures which
show the distributions of the first, the second, and the third JND
points, respectively. Generally speaking, there exhibit large variations
among JND points across different sequences.

\begin{figure}[!thb]
\centering
	\begin{subfigure}[b]{1.0\textwidth}
	\centering
	\includegraphics[width=0.19\linewidth]{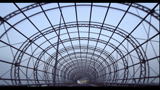}
	\includegraphics[width=0.19\linewidth]{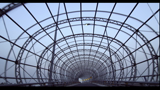}
	\includegraphics[width=0.19\linewidth]{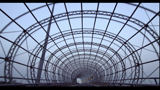}
	\includegraphics[width=0.19\linewidth]{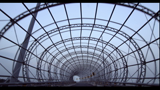}
	\includegraphics[width=0.19\linewidth]{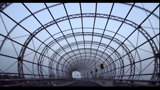}
	\caption{\label{}}
	\end{subfigure}

	\begin{subfigure}[b]{1.0\textwidth}
	\centering{}
	\includegraphics[width=0.19\linewidth]{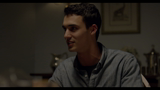}
	\includegraphics[width=0.19\linewidth]{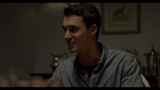}
	\includegraphics[width=0.19\linewidth]{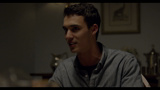}
	\includegraphics[width=0.19\linewidth]{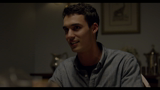}
	\includegraphics[width=0.19\linewidth]{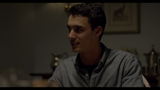}
	\caption{\label{}}
	\end{subfigure}
\caption{Representative frames from source sequences \#15 (a) and \#37
(b), where sequence \# 15 (tunnel) is the scene of a car driving
through a tunnel with the camera mounted on the windshield while source
\# 37 (dinner) is the scene of a dinner table with still camera focusing
on a male speaker.} \label{fig:frames_of_two_sequences}
\end{figure}
We examine sequences \#15 (tunnel) and \#37 (dinner) to offer deeper
insights into the JND distribution. Representative frames are given in
Fig. \ref{fig:frames_of_two_sequences}. Sequence \#15 is a scene with
fast motion and rapid background change. As a result, the masking effect
is strong. It is not a surprise that the JND samples vary a lot among
different subjects. As shown in Fig.
\ref{fig:boxplot_1st_1080p_part_1}, the JND samples of this sequence
have the largest deviation among the 50 sequences in the plot. This
property is clearly revealed by the collected JND samples. Sequence \#37
is a scene captured around a dinner table. It focuses on a male speaker
with dark background. The face of the man offers visual saliency that
attracts the attention of most people. Thus, the quality variation of
this sequence is more noticeable than others and its JND distribution is
more compact. As shown in Fig.  \ref{fig:boxplot_1st_1080p_part_1},
sequence \#37 has the smallest SD among the 50 sequences.

\begin{figure}[!thb]
\centering
\begin{subfigure}[b]{1.0\linewidth}
\centering
\includegraphics[width=0.75\linewidth]{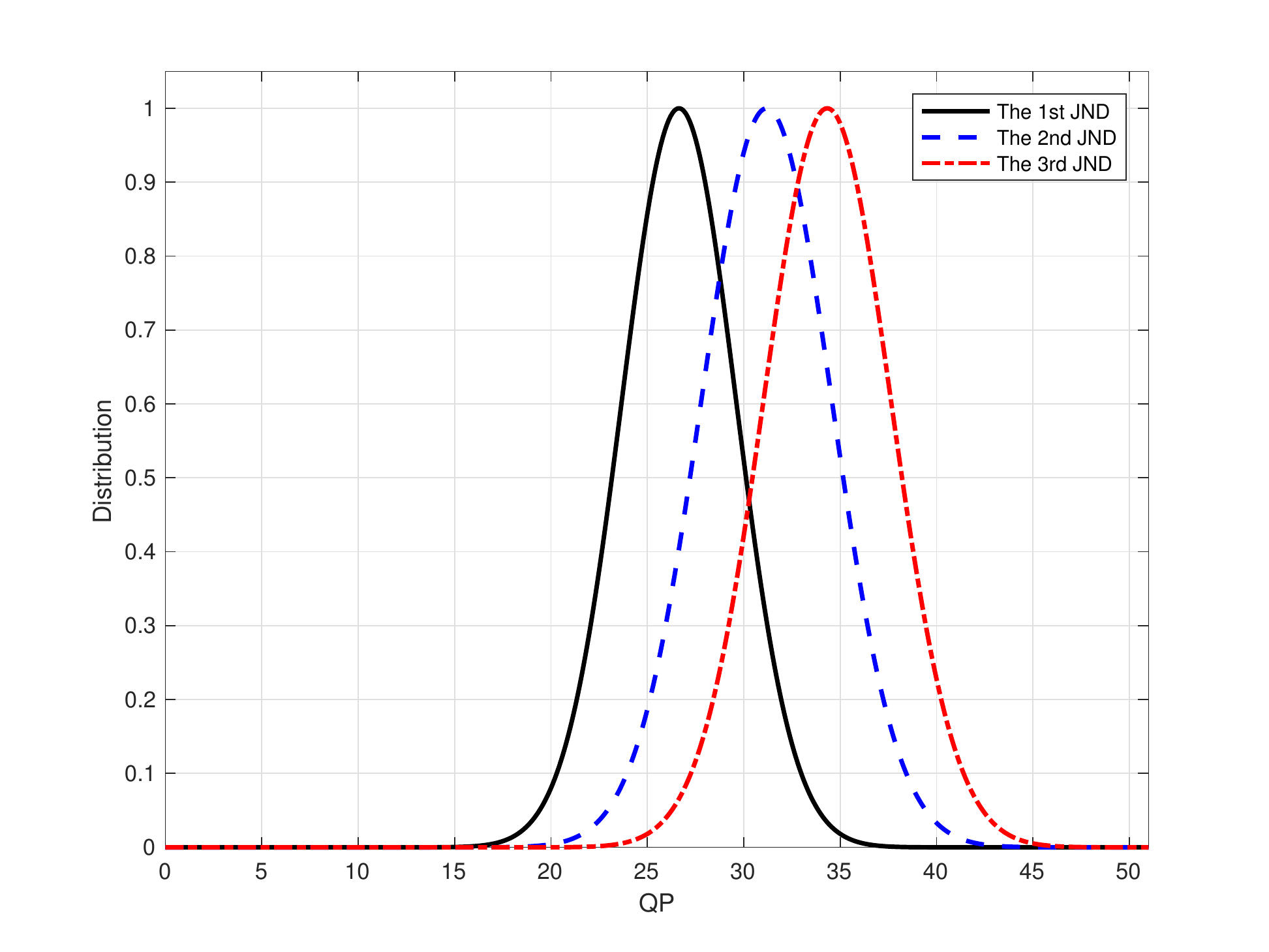}
\phantomcaption
\end{subfigure}
\caption{The histograms of three JND points with all 220 sequences included.\label{fig:3_jnd_is_enough}}
\end{figure}

Furthermore, we plot the histograms of the first, the second, and the
third JND points of all 220 sequences in Fig. \ref{fig:3_jnd_is_enough}.
They are centered around QP = 27, 31 and 34, respectively. For the daily
video service such as the over-the-top (OTT) content, the QP values are
in the range of 18 to 35. Furthermore, take the traditional 5-level
quality criteria as an example (i.e., excellent, good, fair, poor, bad).
The quality of the third JND is between fair and poor. For these
reasons, we argue that it is sufficient to measure 3 JND points. The
quality of coded video clips that go beyond this range is too bad to be
acceptable by today's viewers in practical Internet video streaming
scenarios.

\begin{figure}[!thb]
\centering
	\begin{subfigure}[b]{0.48\linewidth}
	\centering
		\includegraphics[width=1.0\linewidth]{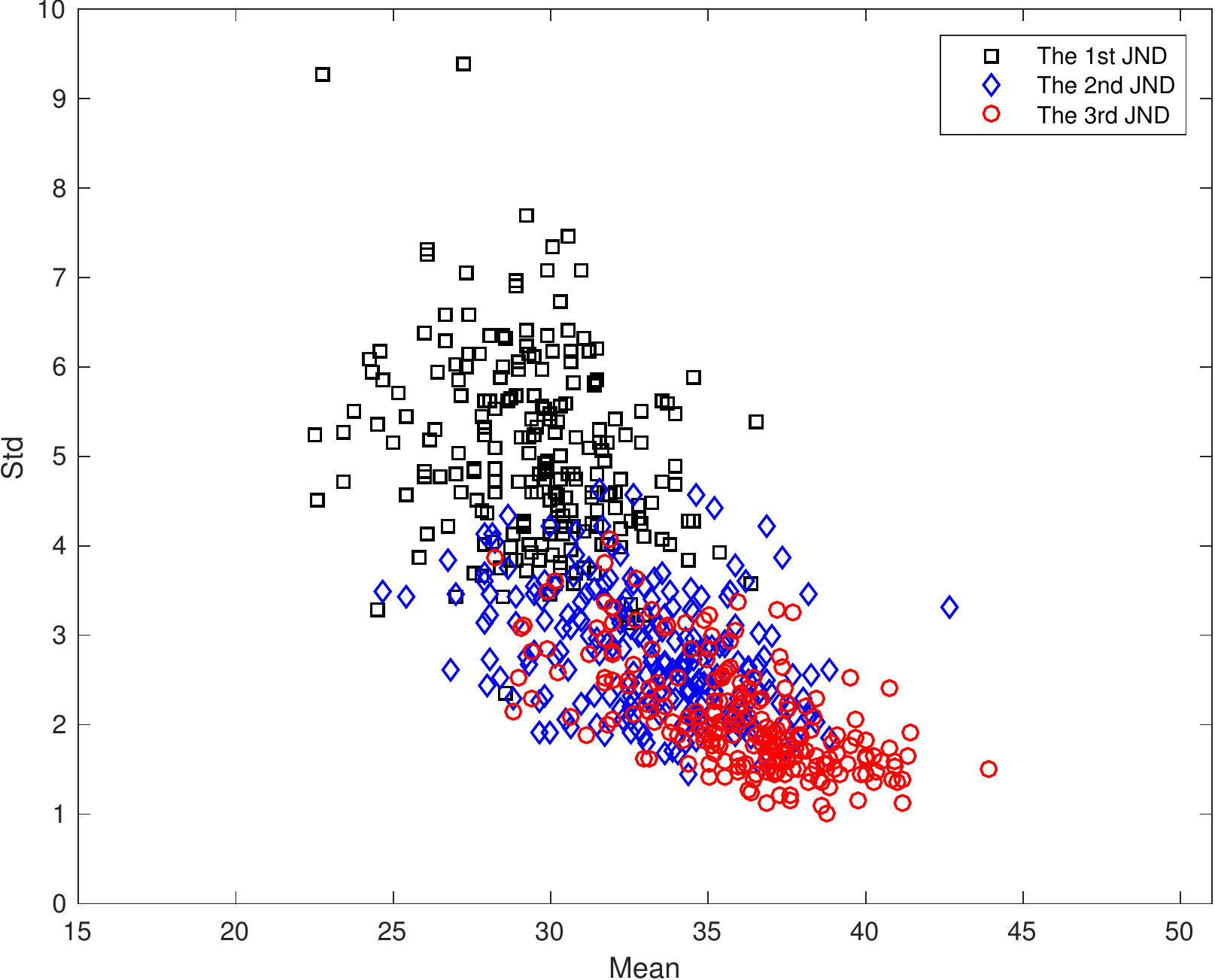}
		\caption{\label{fig:std_mean_fit_1920x1080}}
	\end{subfigure}
	\quad
	\begin{subfigure}[b]{0.48\linewidth}
		\includegraphics[width=1.0\linewidth]{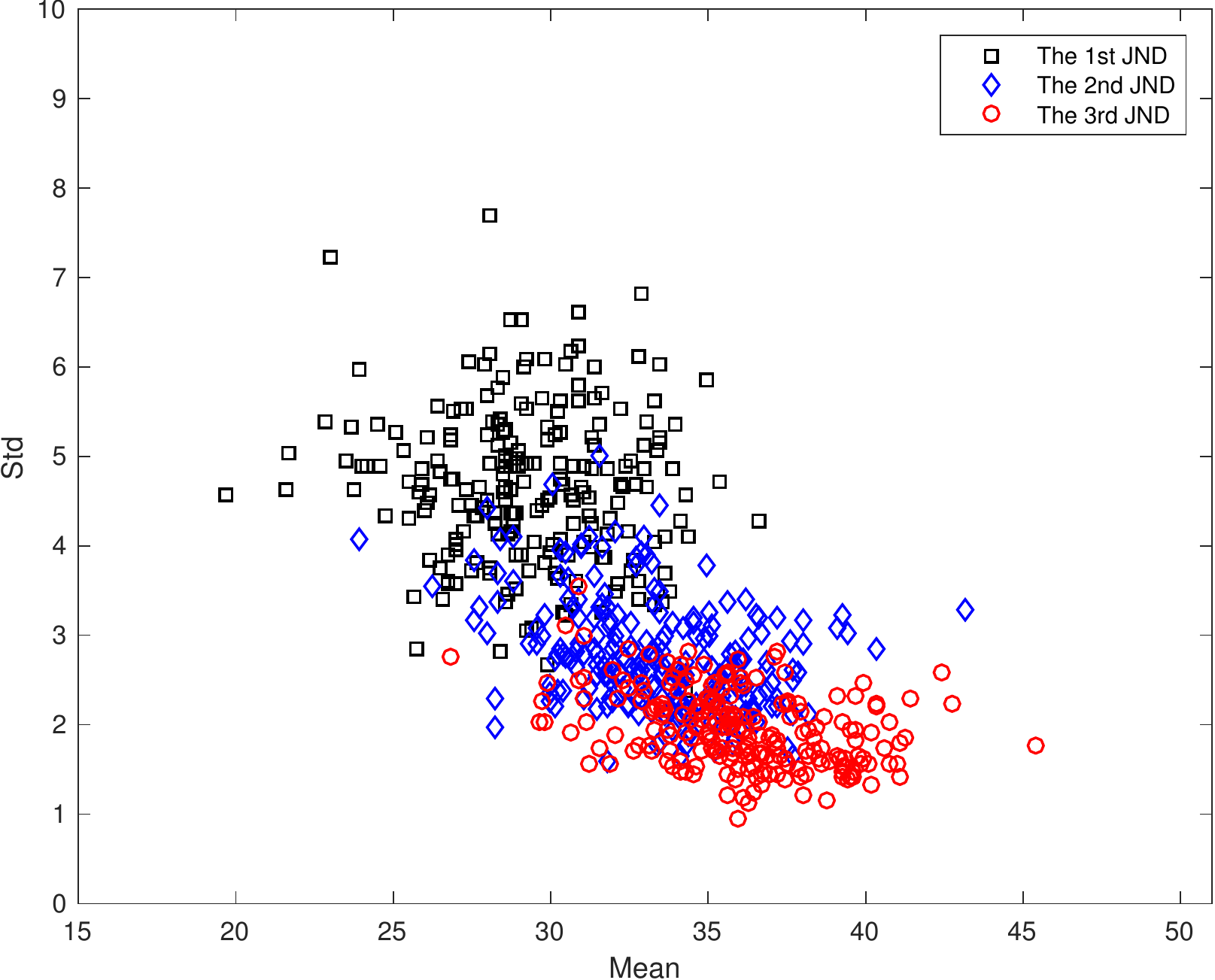}
		\caption{\label{fig:std_mean_fit_1280x720}}
	\end{subfigure}
	\begin{subfigure}[b]{0.48\linewidth}
		\includegraphics[width=1.0\linewidth]{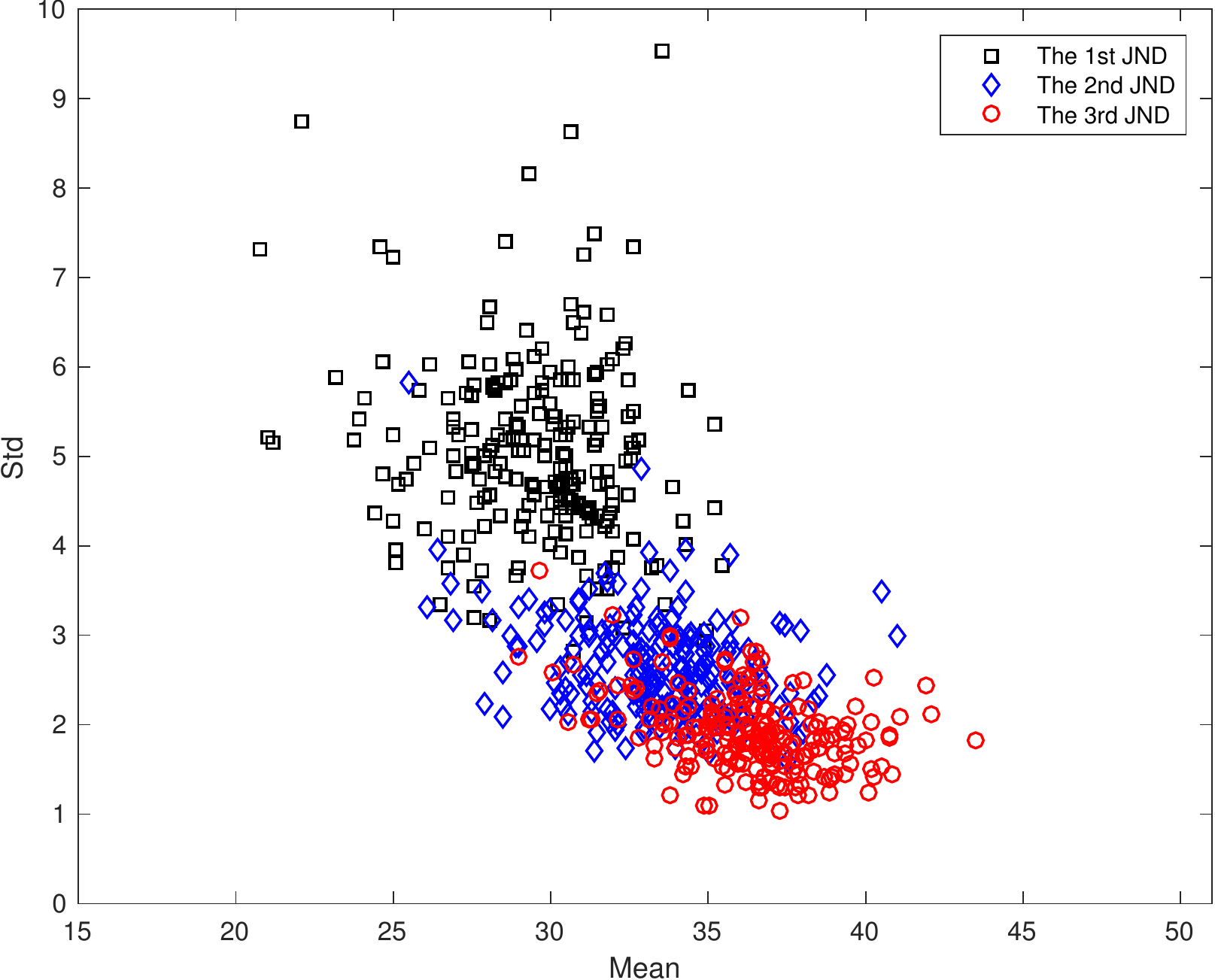}
		\caption{\label{fig:std_mean_fit_960x540}}
	\end{subfigure}
	\quad
	\begin{subfigure}[b]{0.48\linewidth}
		\includegraphics[width=1.0\linewidth]{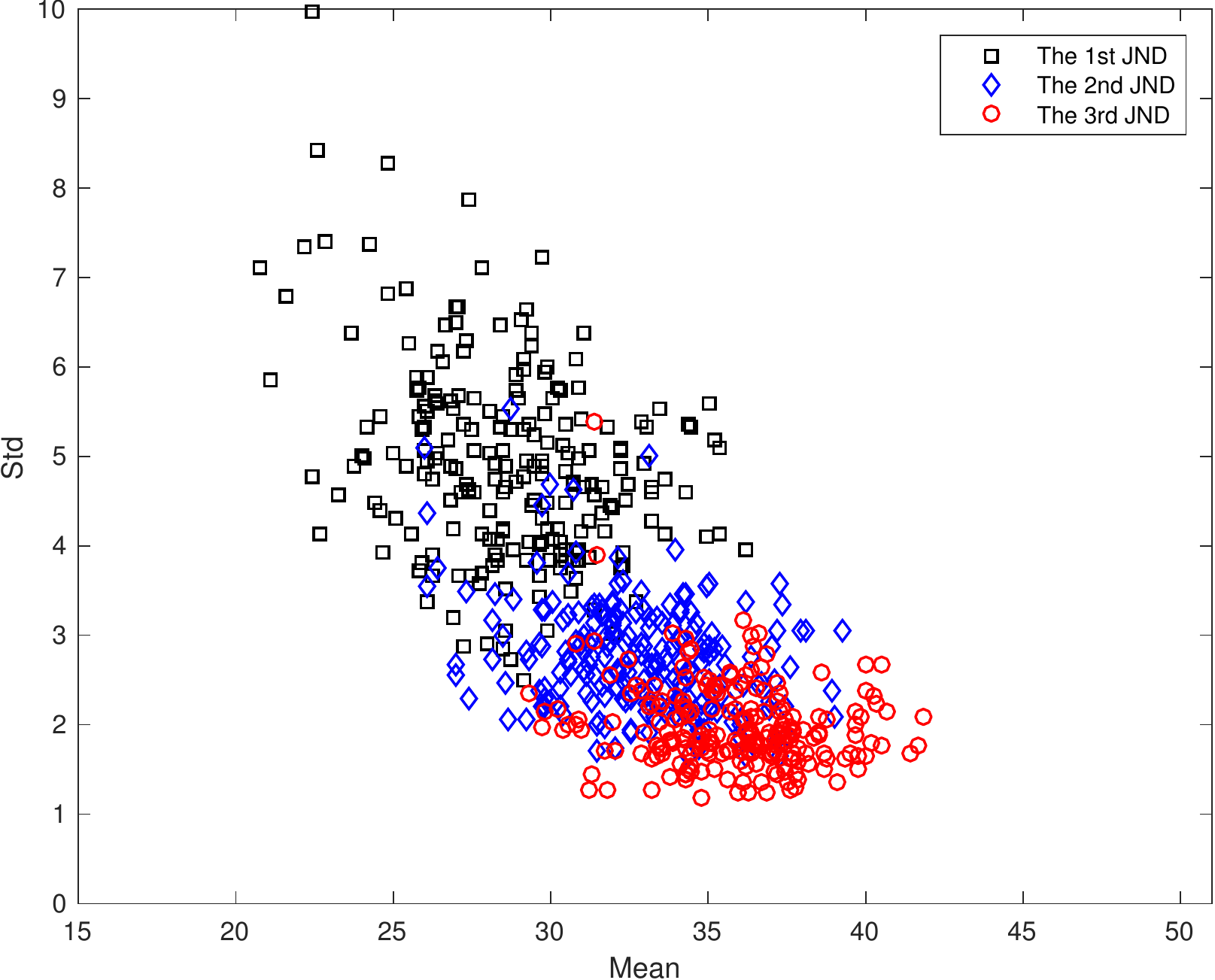}
		\caption{\label{fig:std_mean_fit_640x360}}
	\end{subfigure}
\caption{The scatter plots of the mean/std pairs of JND samples: (a) 1080p, (b) 720p, (c) 540p and (d) 360p.}\label{fig:std_mean_fit}
\end{figure}

The scattered plots of the mean and the SD pairs of JND samples with
four resolutions are shown in Fig. \ref{fig:std_mean_fit}. We observe
similar general trends of the scattered plots in Fig.
\ref{fig:std_mean_fit} in all four resolutions. For example, the SD
values of the second and the third JND points are significantly smaller
than that of the first JND point. The first JND point, which is the
boundary between the perceptually lossy and lossless coded video, is
most difficult for subjects to determine. The main source of observed
artifacts is slight blurriness. In contrast, subjects are more
confident in the decision on the second and the third JND points. The
dominant factor is noticeable blockiness.

The masking effect plays an important role in the visibility of
artifacts. For sequences with a large SD value such as sequence \# 15
in Fig. \ref{fig:boxplot_1st_1080p_part_1}, its masking effect is
strong. On one hand, the JND arrives earlier for some people who are
less affected by the masking effect so that they can see the compression
artifact easily. On the other hand, the compression artifact is masked
with respect to others so that the coding artifact is less visible. For
the same reason, the masking effect is weaker for sequences with a
smaller SD value.

\section{Significance and Implications of VideoSet}\label{sec:significance}

The peak-signal-to-noise (PSNR) value has been used extensively in the
video coding community as the video quality measure. Although it is easy
to measure, it is not exactly correlated with the subjective human
visual experience \cite{Lin-Kuo-2011}. The JND measure demands a great
amount of effort in conducting the subjective evaluation test. However,
once a sufficient amount of data are collected, it is possible to use
the machine learning technique to predict the JND value within a short
interval. The construction of the VideoSet serves for this purpose.

In general, we can convert a set of measured JND samples from a test
sequence to its satisfied user ratio (SUR) curve through integration
from the smallest to the largest JND values. For the discrete case, we
can change the integration operation to the summation operation. For
example, to satisfy $p$\% viewers with respect to the first JND, we can
divide all viewers into two subsets - the first $(100-p)$\% and the
remaining $p$\% - according to ordered JND values. Then, we can set the
boundary QP$_p$ value between the two subsets as the target QP value in
video coding. For the first subset of viewers, their JND value is
smaller than QP$_p$ so that they can see the difference between the
source and coded video clips. For the second subset of viewers, their
JND value is larger than QP$_p$ so that they cannot see the difference
between the source and coded video clips. We call the latter group the
satisfied user group.

\begin{figure}[!t]
\centering
	\begin{subfigure}[b]{1.0\linewidth}
		\centering
		\includegraphics[width=0.75\linewidth]{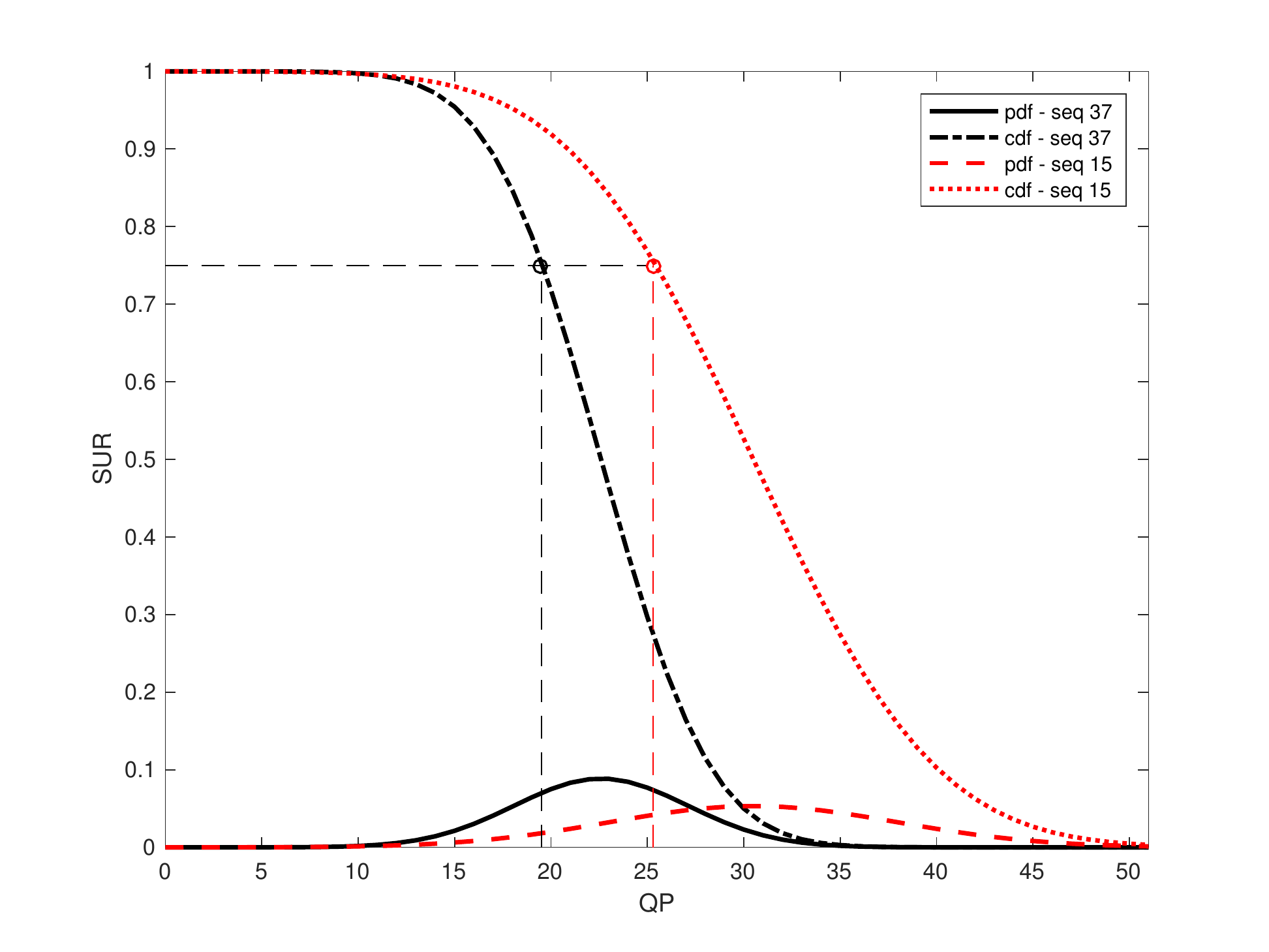}
		\phantomcaption
	\end{subfigure}
\caption{The JND and the SUR plots of sequence \#15 ($\mu=30.5, \sigma=7.5$)
and sequence \#37 ($\mu=22.6, \sigma=4.5$). \label{fig:SUR}}
\end{figure}

When we model the JND distribution as a normal distribution, the SUR
curve becomes the Q-function. Two examples are given in Fig.
\ref{fig:SUR}, where the first JND points of sequence \#15 and \#37 are
plotted based on their approximating normal distributions, where the
mean and SD values are derived from the subjective test data. Their
corresponding Q-functions are also plotted. The Q-function is the same
as the SUR curve. For example, the top quartile of the Q-function gives
the QP value to encode the video content whose quality will satisfy 75\%
of viewers in the sense that they cannot see the difference between the
coded video and the source video. In other words, it is perceptually
lossless compression for these viewers.

\begin{figure}[!thb]
	\centering
	\begin{subfigure}[b]{0.24\linewidth}
		\centering
		\includegraphics[width=1.0\linewidth]{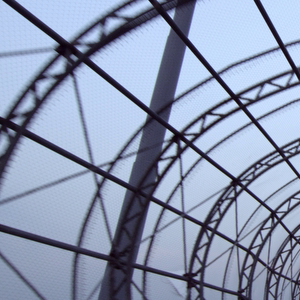}
		\caption{}
	\end{subfigure}
	\begin{subfigure}[b]{0.24\linewidth}
		\centering
		\includegraphics[width=1.0\linewidth]{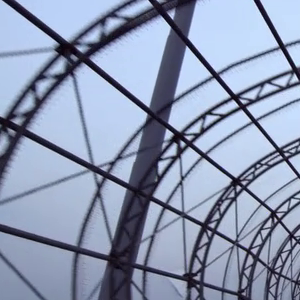}
		\caption{}
	\end{subfigure}
	\begin{subfigure}[b]{0.24\linewidth}
		\centering
		\includegraphics[width=1.0\linewidth]{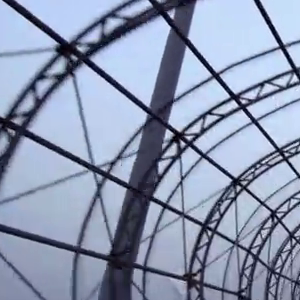}
		\caption{}
	\end{subfigure}
	\begin{subfigure}[b]{0.24\linewidth}
		\centering
		\includegraphics[width=1.0\linewidth]{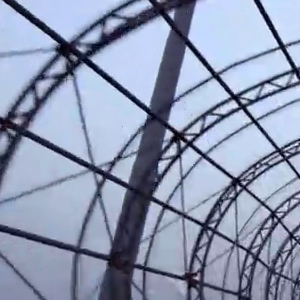}
		\caption{}
	\end{subfigure}
	\begin{subfigure}[b]{0.24\linewidth}
		\centering
		\includegraphics[width=1.0\linewidth]{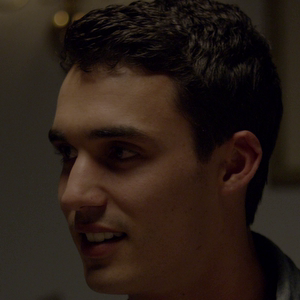}
		\caption{}
	\end{subfigure}
	\begin{subfigure}[b]{0.24\linewidth}
		\centering
		\includegraphics[width=1.0\linewidth]{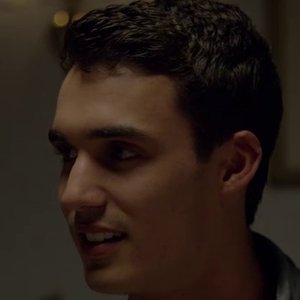}
		\caption{}
	\end{subfigure}
	\begin{subfigure}[b]{0.24\linewidth}
		\centering
		\includegraphics[width=1.0\linewidth]{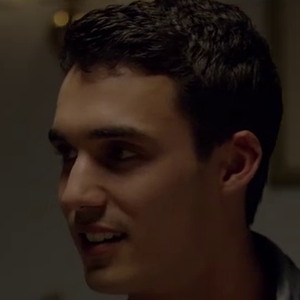}
		\caption{}
	\end{subfigure}
	\begin{subfigure}[b]{0.24\linewidth}
		\centering
		\includegraphics[width=1.0\linewidth]{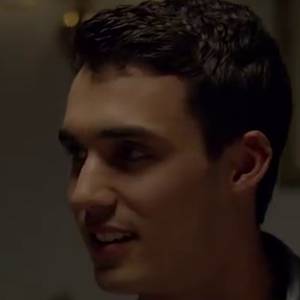}
		\caption{}
	\end{subfigure}
\caption{Comparison of perceptual quality of two coded video sequences.
Top row: (a) the reference frame with QP=0, (b) the coded frame with QP=25,
(c) the coded frame with QP=36, and (d) the coded frame with QP=38,
of sequence \#15.
Bottom row: (e) the reference frame with QP=0, (f) the coded frame with QP=19,
(g) the coded frame with QP=22, and (h) the coded frame with QP=27,
of sequence \#37.\label{fig:zoomed_frames}}
\end{figure}

We show four representative thumbnail images from the two examples in
Fig. \ref{fig:zoomed_frames}. The top and bottom rows are encoded
results of sequence \#15 and sequence \#37, respectively. The first
column has the best quality with QP=0. Columns 2-4 are encoded with the
QP values of the first quartiles of the first, the second, and the third
JND points. For a great majority of viewers (say, 75\%), the video clip
of the first JND point is perceptually lossless to the reference one as
shown in the first column. The video clip at the second JND point begins
to exhibit noticeable artifacts. The quality of the video clip at the
third JND point is significantly worse.

The VideoSet and the SUR quality metric have the following four important
implications.
\begin{enumerate}
\item It is well known that the comparison of PSNR values of coded video
of different contents does not make much sense. In contrast, we can
compare the SUR value of coded video of different contents. In other
words, the SUR value offers a {\em universal} quality metric.
\item We are not able to tell whether a certain PSNR value is sufficient
for some video contents. It is determined by an empirical rule. In
contrast, we can determine the proper QP value to satisfy a certain
percentage of targeted viewers. It provides a practical and
theoretically solid foundation in selecting the operating QP for rate
control.
\item To the best of our knowledge, the VideoSet is the largest scale subject
test ever conducted to measure the response of the human visual system
(HVS) to coded video. It goes beyond the PSNR quality metric and opens a
new door for video coding research and standardization, {\em i.e.}
data-driven perceptual coding.
\item Based on the SUR curve, we can find out the reason for the existence of the first JND point. Then, we can try to mask the
noticeable artifacts with novel methods so as to shift the first JND
point to a larger QP value. It could be easier to fool human eyes than
to improve the PSNR value.
\end{enumerate}

\section{Conclusion and Future Work}\label{sec:conclusion}

The construction of a large-scale compressed video quality dataset based
on the JND measurement, called the VideoSet, was described in detail in
this paper. The subjective test procedure, detection and removal of
outlying measured data, and the properties of collected JND data were
detailed. The significance and implications of the VideoSet to future
video coding research and standardization efforts were presented. It
points out a clear path to data-driven perceptual coding.

One of the follow-up tasks is to determine the relationship between the
JND point location and the video content. We need to predict the mean
and the variance of the first, second and third JND points based on the
calibrated dataset; namely, the VideoSet. The application of the machine
learning techniques to the VideoSet for accurate and efficient JND
prediction over a short time interval is challenging but an essential
step to make data-driven perceptual coding practical for real world
applications. Another follow-up task is to find out the artifacts
caused by today's coding technology, to which humans are sensitive. Once
we know the reason, it is possible to mask the artifacts with some novel
methods so that the first JND point can be shifted to a larger QP value.
The perceptual coder can achieve an even higher coding gain if we take
this into account in the next generation video coding standard.

\section*{Acknowledgments}

This research was funded by Netflix, Huawei, Samsung and MediaTek. The
subjective tests were conducted in the City University of Hong Kong and
five universities in the Shenzhen City of China. They were Shenzhen
University, Chinese University of Hong Kong (SZ), Tsinghua University,
Peking University and Chinese Academy of Sciences. Computation for the
work was supported in part by the University of Southern California's
Center for High-Performance Computing (hpc.usc.edu). The authors would
like to give thanks to these companies and universities for their strong
support.


\bibliographystyle{splncs}
\bibliography{refs}

\begin{thebibliography}{10}

\bibitem{lin2015experimental}
Lin, J.Y., Jin, L., Hu, S., Katsavounidis, I., Li, Z., Aaron, A., Kuo, C.C.J.:
\newblock Experimental design and analysis of {JND} test on coded image/video.
\newblock In: SPIE Optical Engineering+ Applications, International Society for
  Optics and Photonics (2015)  95990Z--95990Z

\bibitem{jin2016jndhvei}
Jin, L., Lin, J.Y., Hu, S., Wang, H., Wang, P., Katasvounidis, I., Aaron, A.,
  Kuo, C.C.J.:
\newblock Statistical study on perceived {JPEG} image quality via {MCL-JCI}
  dataset construction and analysis.
\newblock In: IS\&T/SPIE Electronic Imaging, International Society for Optics
  and Photonics (2016)

\bibitem{mcl_jcv}
Wang, H., Gan, W., Hu, S., Lin, J.Y., Jin, L., Song, L., Wang, P.,
  Katsavounidis, I., Aaron, A., Kuo, C.C.J.:
\newblock {MCL-JCV}: A {JND}-based {H.264/AVC} video quality assessment
  dataset.
\newblock In: 2016 IEEE International Conference on Image Processing (ICIP).
  (2016)  1509--1513

\bibitem{h2h01c-16-full}
Haiqiang, W., Ioannis, K., Xin, Z., Jiwu, H., Man-On, P., Xin, J., Ronggang,
  W., Xu, W., Yun, Z., Jeonghoon, P., Jiantong, Z., Shawmin, L., Sam, K., Kuo,
  C.C.J.:
\newblock Videoset: A large-scale compressed video quality dataset based on
  {JND} measurement.
\newblock \url{https://ieee-dataport.org/documents/videoset} (2016)

\bibitem{aimar2005x264}
Aimar, L., Merritt, L., Petit, E., Chen, M., Clay, J., Rullgrd, M., Heine, C.,
  Izvorski, A.:
\newblock X264-a free h264/avc encoder.
\newblock \url{http://www.videolan.org/developers/x264.html} (2005) Accessed:
  04/01/07.

\bibitem{CDVL}
Pinson, M.H.:
\newblock The consumer digital video library [best of the web].
\newblock \url{http://www.cdvl.org/resources/index.php} (2013)

\bibitem{Cablelabs}
CableLabs.
\newblock (\url{http://www.cablelabs.com/resources/4k/}) Accessed: 2016-11-22.

\bibitem{tos}
{B}lender {F}oundation, C.
\newblock (\url{mango.blender.org}) Accessed: 2016-11-22.

\bibitem{ou2014q}
Ou, Y.F., Xue, Y., Wang, Y.:
\newblock Q-star: a perceptual video quality model considering impact of
  spatial, temporal, and amplitude resolutions.
\newblock IEEE Transactions on Image Processing \textbf{23} (2014)  2473--2486

\bibitem{lanczos}
Turkowski, K.:
\newblock Graphics gems.
\newblock Academic Press Professional, Inc., San Diego, CA, USA (1990)
  147--165

\bibitem{sharma2002digital}
Sharma, G., Bala, R.:
\newblock Digital color imaging handbook.
\newblock CRC press (2002)

\bibitem{grubbs1950sample}
Grubbs, F.E.:
\newblock Sample criteria for testing outlying observations.
\newblock The Annals of Mathematical Statistics (1950)  27--58

\bibitem{jarque1987test}
Jarque, C.M., Bera, A.K.:
\newblock A test for normality of observations and regression residuals.
\newblock International Statistical Review/Revue Internationale de Statistique
  (1987)  163--172

\bibitem{Lin-Kuo-2011}
Lin, W., Jay~Kuo, C.C.:
\newblock Perceptual visual quality metrics: A survey.
\newblock Journal of Visual Communication and Image Representation \textbf{22}
  (2011)  297--312

\end{thebibliography}

\end{document}